\documentclass[12pt,preprint,isolatin1,graphicx]{aastex}
\shorttitle{The central arcsecond of our galaxy}
\shortauthors{Sch\"odel et al.}

\received{2003 February 21}
\begin{document}


\title{Stellar dynamics in the central arcsecond of our
  galaxy\footnote{Based on observations at the Very Large Telescope
  (VLT) of the European Southern Observatory (ESO) on Paranal in
  Chile}}


\author{R. Sch\"odel, T. Ott, R. Genzel\altaffilmark{1}}
\affil{Max-Planck-Institut f\"ur extraterrestrische Physik,
  Giessenbachstra\ss e, Garching, Germany}
\altaffiltext{1}{Also: Department of Physics, University of California,
    Berkeley, CA 94720}

\and

\author{A. Eckart, N. Mouawad}
\affil{I.Physikalisches Institut, Universit\"at zu K\"oln, Z\"ulpicher
  Stra\ss e , K\"oln, Germany}

\and

\author{T. Alexander}
\affil{Faculty of Physics, Weizmann Institute of Science, Rehovot, Israel}

%%ABSTRACT

\begin{abstract}
With 10 years of high-resolution imaging data now available on the
stellar cluster in the Galactic Center, we present proper motions for
$>$40 stars at projected distances $\leq1.2''$ from Sagittarius A*
(Sgr~A*).  We find evidence on a $\geq2\sigma$ level for radial
anisotropy of the cluster of stars within $1''$ of Sgr~A*. For a
brightness limit of $\mathrm{K}\sim15.5$ we find no evidence for a
stationary source at the position of Sgr~A* or for a source at this
position that would be variable on a time scale of at least several
hours to days. On time scales of seconds to tens of minutes, we find
no variability at the Sgr~A* position on brightness levels
$\mathrm{K}\leq13.5$. We confirm/find accelerated motion for 6 stars,
with 4 stars having passed the pericenter of their orbits during the
observed time span. We calculated/constrained the orbital parameters
of these stars. All orbits have moderate to high eccentricities. We
discuss the possible bias in detecting preferentially orbits with high
eccentricities and find that measured values of $e>0.9$ might be
detected by about a factor of $1.5-2$ more frequently.  We find that
the center of acceleration for all the orbits coincides with the radio
position of Sgr~A*. From the orbit of the star S2, the currently most
tightly constrained one, we determine the mass of Sgr~A* to be
$3.3\pm0.7\times10^{6}$M$_{\odot}$ and its position to $2.0\pm2.4$~mas
East and $2.7\pm4.5$~mas South of the nominal radio position. The mass
estimate for the central dark mass from the orbit of S2 is fully
consistent with the mass estimate of
$3.4\pm0.5\times10^{6}$M$_{\odot}$ obtained from stellar proper
motions within $1.2''$ of Sgr~A* using a Leonard-Merritt mass
estimator. We find that radio astronomical observations of the
proper motion of Sgr~A* in combination with its intrinsic
source size place at the moment the tightest constraints on the mass
density of Sgr~A*, which must exceed
$\rho_{\mathrm{Sgr~A*}}>3\times10^{19}\mathrm{M}_{\odot}\mathrm{pc}^{-3}$.
\end{abstract}



%%MAIN BODY

\section{Introduction}

The discovery of quasars and active galactic nuclei gave rise to the
idea that massive black holes (MBH) of several million to billion
solar masses exist at the centers of active galaxies. Over the last
decades, evidence has been accumulated through the observation of
stellar and gas dynamics in the centers of nearby galaxies, that most
quiet galaxies harbour a MBH at their centers as well, with more than
30 candidates listed in the latest census \citep{kormendy01} \footnote{an
up-to-date list is maintained at {\em
http://chandra.as.utexas.edu/$\sim$kormendy/bhsearch.html}}. With the
number of putative MBH growing constantly through increasing observational
evidence, it has meanwhile become a standard paradigm of modern astronomy
that MBHs exist at the centers of the vast majority of galaxies.

A special case is the center of our home galaxy. Located at a distance
of only 8~kpc from the solar system \citep{reid93}, it allows detailed
astrometric observations of stars at distances $\ll1$pc from the
central black hole candidate, the radio source Sgr~A*. Near-Infrared
(NIR) high-resolution imaging and spectroscopic observations with
speckle and adaptive optics techniques were carried out during the
last decade at 4m to 10m-class telescopes. They allowed to measure the
projected and radial motions of $\sim$100 stars within the central
0.1~pc as well as the accelerations of a few individual stars as close
as a few light days from Sgr~A*. The results of these observations
made a compelling case that the concentration of dark mass seen at the
center of the Milky Way is indeed in the form of a supermassive black
hole \citep{genzel96, eckart97, ghez98, ghez00, eckart02, schoedel02}.

A series of observations that covered the pericenter passage of the
star S2 around Sgr~A*, allowed \citet{schoedel02} to determine a
unique keplerian orbit (up to the sign of the inclination) for this
object and to measure the enclosed dark mass down to a distance of a
mere 17 light hours from Sgr~A*. With these observations, they could
exclude with high confidence a neutrino ball scenario
\citep{munviol02} as an alternative explanation for the dark mass
concentration as well as a cluster of dark astrophysical objects
\citep{maoz98}, such as neutron stars, leaving a central supermassive
black hole as the only plausible explanation. Recently, \citet{ghez03}
confirmed the orbit of S2, presenting also the first line-of-sight
velocity measurements for this star. 

With the nature of the dark central object now being known to a high
degree of confidence, we want to understand the interactions of the
black hole with its environment and its impact on the population of
surrounding stars and their dynamics. In this paper we provide a
high-quality data base of proper motions and accelerations of the
stars in the immediate environment of Sgr~A*, on scales of $\sim1''$,
or 0.04~pc (for a Galactic Center distance of 8~kpc \citep{reid93},
which will be assumed throughout this paper). With these data we
analyze the dynamics of this region which is dominated by the stellar
cusp around Sgr~A* that was found in deep adaptive optics (AO)
observations by \citet{genzel03}.

\section{Observations and data reduction \label{data}}

We compiled observations from three different data sets for the
present work. Near-infrared (NIR) speckle imaging observations with
the MPE-built NIR speckle imaging camera SHARP \citep{hofmann95} were
carried out at the ESO 3.5m NTT in La Silla, Chile, from 1992 to
2002. For the year 2000 we included the publicly available Gemini
North Observatory Galactic Center Demonstration Science Data Set.  For
the 2002 epoch we used the commissioning and science verification AO
observations of the GC with the new NAOS/CONICA adaptive optics
system/NIR camera (constructed by MPIA/MPE, Meudon/Grenoble
Observatories, ONERA, ESO) on the ESO
VLT unit telescope 4 (Yepun) on El Paranal, Chile \citep{rousset98,
lenzen98}. The latter data are publicly available in the ESO archive
and some of the first results have been published by
\citet{schoedel02} and \citet{genzel03}. The individual data sets and
their processing are described in more detail in the following
sections. Standard data reduction procedures, i.e. sky subtraction,
dead/bad pixel masking, and flat-fielding, were applied to all the
imaging data.

\subsection{The SHARP/NTT imaging data \label{sharpdata}}

Imaging data with the MPE speckle camera SHARP were obtained once or
twice every year since 1992, with the instrument used as a guest
instrument at the ESO NTT. An integration time of 0.5~s was used for
the individual imaging frames. This was found to be the best choice
for achieving an optimal signal-to-noise ratio while at the same time
still conserving the diffraction limited information. From the several
tens of thousands of individual speckle frames obained in each
observing epoch, we pre-selected $\sim$1000 high-quality frames
automatically. Selection criterion was the number of bright speckles
of the dominant star in a frame: Beside the brightest speckle, there
should not be more than one additional speckle with more than 10\% of
the flux of the brightest speckle. From the pre-selected frames we
picked by eye several hundred speckle frames with the highest
quality. Selection criterion was that the first diffraction ring
around the dominant speckle of the brightest stars must be clearly
visible in these individual 0.5~s exposures. For every observing
epoch, the final selection of highest quality images was devided into
$\sim4$ subsets of about one hundred frames each.  After combining the
frames of such a subset with a simple shift-and-add algorithm, the
Strehl ratio was found to be of the order 30\% in the final images.

We applied Iterative Blind Deconvolution (IBD) \citep{jeffries93} on
the sets of high-quality speckle frames. IBD iteratively reconstructs
the object and the point spread function of an image from initial
estimates. It makes use of some general constraints on the imaging
properties, e.g. the positivity of object and PSF and that the
convolution of the object with the PSF must yield the original image
again (in case of zero noise). In the case of speckle imaging, we have
multiple images of the same object. IBD can take use of that
information effectively. It reconstructs the observed object and at
the same time the individual PSFs for each speckle frame. Convergence
of the IBD algorithm is guided via an error metric that describes the
deviation of the PSFs and object estimates from their constraints.

We used the publicly available IDAC program code \footnote{developed at Steward
Observatory by Matt Chesalka and Keith Hege (based on the earlier
Fortran Blind Deconvolution code - IDA - developed by Stuart Jefferies
and Julian Christou)}.  The shift-and-add images of the sets of speckle
frames were used for the initial estimates. Because of the high demand
of computing time for this algorithm, we applied it only to a
$\sim6\times6''$ area around Sgr~A* that included the IRS 16 complex
of bright (K$\sim$10) sources. After convergence of the IBD algorithm,
the resulting 3 to 4 object images for every observing epoch were
oversampled to a third of the original pixel scale, smoothed and
coadded.  Any remaining PSF residuals were then cleaned with a
Lucy-Richardson deconvolution. We produced final maps by restoring the
result of the deconvolution with a Gaussian beam of 100 mas FWHM. 

We also produced maps in a similar way by using the Lucy-Richardson
(LR) algorithm instead of the IDAC code and the shift-and-add images
instead of the individual speckle frames. The maps compare very well
with the maps produced with the aid of the IDAC code. However, we
found that the IDAC procedure provided maps of practically constant
quality for all epochs, while the quality is slightly more variable in
the case of the LR algorithm. A possible cause for this may be that
one needs to use a PSF estimate for a LR deconvolution. We used the
bright, isolated source IRS 7 as PSF reference. It is $\sim5$
magnitudes brighter than any other star in its immediate surroundings
and is thus very well suited as a PSF reference. However, in some
cases the PSF reference was not contained in the FOV, in other cases
the anisoplanatic angle might have been small. Therefore the results
of the LR deconvolved maps are of slightly more variable quality than
the IBD maps and we decided to use the latter in our further analysis
because they guarantee an equal treatment and quality of all the maps.

One of the strengths of the IBD/IDAC routine is that it can use {\em
multiple} images of the {\em same} object at the same time, which
poses tight constraints on the object distribution. Thus, the final
object can be reconstructed reliably. A further advantage of IBD is
that one does not need an accurate estimate of the PSF for a reliable
deconvolution. On the negative side, one has to mention the
considerable amount of computing time needed for the IBD algorithm.

In the described way, we obtained high-resolution maps for all epochs,
except for 1993 and 2002, where the quality of the imaging data did
not allow to select a sufficiently large sample of high-quality
speckle frames. The 1992 speckle frames are characterised by a fairly
strong readout ``waffle'' pattern. While the resulting high-resolution
map for 1992 compares favourably with the other epochs, the stellar
positions may be subject to larger errors for this epoch.

Our method here is optimized for producing the deepest
(K(3-5$\sigma$)$\approx$16) diffraction limited SHARP/NTT maps of comparable
quality for all observing epochs. This is in contrast to our earlier
work, such as \citet{eckart97}. There, deep high-resolution maps were
produced only for a few selected epochs. The approach of \citet{ott03}
is also distinct from the approach presented here: They produced and
analyzed in a largely automated, efficient way several tens of
shift-and-add images for every SHARP observing period. While their
maps allow to effectively determine proper motions for $\sim$1000
stars in a $\sim10''$ radius region around Sgr~A*, they are generally
$~\sim$0.5 to 1 magnitude shallower and of more variable quality than
the maps used in this work, which serve the purpose to examine the
proper motions in the very central, most crowded part of the cluster,
and for the weakest sources possible. The \citet{ott03} aproach and our
approach presented here can be seen as mutually complementary.

In order to verify the SHARP high-resolution maps, we compared them
among each other to check whether the identified sources and proper
motions were consistent between the epochs (see
section~\ref{positions}). An additional check was made by a comparison
between the IBD and the LR deconvolved maps. For the epochs around
2000, we also compared the SHARP maps with the maps obtained from the
Gemini and NACO data sets (see below). In Figure~\ref{comparison} we
show the the Gemini 2000, the SHARP 2001 (IBD and LR), and NACO 2002
maps, all of them restored to a FWHM of $\sim$100~mas.

\subsection{The Gemini imaging data}

In addition to our SHARP map for the epoch 2000, we selected one
K'-band image from the publicly available Gemini North Observatory
Galactic Center Demonstration Science Data Set. This data set provides
observations of the Galactic Center stellar cluster with the Gemini
North telescope, the Quirc near infrared camera and the Hokupa'a AO
system. We selected an image of 750~s total integration time, oberved
in July 2000. Since the Hokupa'a AO system was designed for a 4m-class
telescope, it could only partially provide the correction needed for
the 8m-class Gemini telescope. Moreover, the visible guiding star used
for the observations is located at $\sim30''$ from Sgr~A*, so the
Strehl ratio of the image is rather low. However, the central sources
around Sgr~A* are clearly resolved at an estimated resolution of
100~mas.

We created a PSF from the median of $\sim15$ bright, unsaturated stars
in the image. Because the wings of the PSF were very strong, but
contaminated by weak sources, we fitted the wings of the PSF with a
Moffat function. Subsequently, we created a model PSF by combining the
model for the wings with the PSF Kernel obtained by the median of the
stellar images. This PSF model was used for a LR deconvolution of the
K'-band image, which was reconvolved to a final map with a Gaussian
beam of $\sim$100 mas FWHM. We did not use the IDAC algorithm because
it works less well in case there is only one image of a given
object.

\subsection{The NAOS/CONICA imaging data}

The GC stellar cluster was observed several times during the
commissioning and science verification of the NAOS/CONICA (``NACO'')
AO system and near infrared camera at the ESO VLT unit telescope
4. With its unique infrared wavefront sensor, the loop of the AO was
closed on the K$\sim6.5$~mag supergiant IRS~7, which is located
$\sim5.5''$ north of Sgr~A*. We chose K'-band observations from May
2002 (30 frames with 30~s integration time) and from August 2002 (20
frames with 60~s integration time). The individual frames were
coadded to images with a total of 900~s (May) and 1200~s (August)
integration time. In the August observation run, a Strehl ratio of $>
50\%$ was achieved. A description of the NACO imaging data can also be
found in \citet{genzel03} and \citet{ott03a}.

We used $\sim$15 stars in each image to extract a median PSF, which was
used for a Lucy-Richardson deconvolution. Final maps were obtained
after restoration with a Gaussian beam of $\sim$60~mas FWHM. As in the
case of the Gemini image, LR deconvolution was used because of its
simpler use and greater speed. Since the NACO images are of much
better quality (resolution and deepness) than the Gemini/SHARP data
sets, deconvolution was not a critical issue in their case.

\section{Positions and proper motions\label{positions}}

The elaborate processing of the imaging data provided us with 12 final
maps with resolutions between 60 to 130 mas FWHM for the epochs
between 1992 and 2002 (no map for 1993, two maps for the 1996, 2000,
and 2002 epochs each). These maps show the global evolution of the
Sgr~A* central cluster. Source confusion in this cluster is high, so
that because of the high proper motions a star may ``merge'' at some
epoch with another one and ``reappear'' at some later epoch. With a
baseline that is long enough, it is possible to disentangle these
stellar motions. Source identification (and reidentification at a
later epoch) is usually done ``by eye''. In order to cross-check and
support our ``by eye'' identifications, we implemented a largely
automatized procedure that, given initial source positions and
identifications at one epoch, identifies the selected sources and
measures their positions in the maps of all other epochs.

Assuming constant relative fluxes of the stars, the appearance of the
stellar cluster at different epochs results just from rearrangements
of the stars. If the differences in the positions of the stars are not
too large, which is e.g. the case for two maps from subsequent
observing epochs, this rearrangement can be done automatically with a
least squares fit (that also takes into account an offset in the
overall flux and adapts the PSF FWHM to the respective images). To
start the position finding algorithm, we measured the positions and
relative fluxes of stars in the cluster in the NACO August 2002 image,
the deepest and highest resolution image in our data set. From these
initial estimates, a model image (using Gaussian PSFs) was created
that was subsequently fit (in a least square sense) to the maps of the
stellar cluster at the earlier epochs. Hence, the measured positions
of the stars at one observing epoch were used as initial estimates for
the positions in the preceding epoch. 9 bright, isolated reference
stars served for transforming the initial positions into the
respective individual image frames with their unique rotation angles
and different pixel scales.

We implemented the procedure described above in a series of IDL
program codes and extracted the stellar positions in three
steps. First, the brighter members of the cluster around Sgr~A* were
fitted and subtracted from the maps.  The procedure was then repeated
with fainter stars on the subtracted maps from step 1. In a third
iteration final positions were obtained by taking the positions of the
brighter and fainter stars measured in steps 1 and 2 for all epochs
and fitting them simultaneously to the original maps of all
epochs. With this final iteration, we tried to minimize the influence
that sources  very close to each other have on their mutual
positions.  Errors on the measured positions were determined by
comparing the positions from the fitting procedure with centroid positions
measured on the stars with two different apertures sizes and taking
the maximum deviation of the centroid positions from the fitted
position as an error estimate.

With the positions (in pixels) measured by the procedure described
above, the offsets of the stars from Sgr~A* in right ascension and
declination were calculated by a second order transformation onto an
astrometric frame using the positions and proper motions of 9 bright,
isolated stars in our field of view from the \citet{ott03} list. This
means that our astrometric system is ultimately established by the SiO
maser astrometry defined by \citet{reid03}. Errors of this
transformation were estimated by taking the standard deviations of 9
measured positions, which we obtained by repeating the
transformation with different subsets of 8 out of the nine stars. The
transformation errors were added quadratically to the positional
errors described above.

We determined stellar proper motion velocities by a linear least
squares fit to the time dependent stellar positions. In a final step,
we controlled the derived positions and proper motions by comparing
model images created with the measured quantities to the observed
images at each observing epoch. In a few cases, the automated
procedure failed to disentangle the motions of stars that were at
coincident positions at some epochs and of comparable magnitudes. In
some other cases, very faint stars ``merged'' with bright stars at
some epoch, but it could not be determined reliably when and where
they ``reappeared''. We rejected these sources with ambiguous proper
motions from our list.

Based on the entire data set, we thus determined proper motions for 35
stars within $1.2''$ of Sgr~A*. The main limitation to our sample
comes from the resolution of the SHARP/NTT data. We could measure the
proper motion of an additional 11 stars based on the higher resolution
Gemini 2000 and NACO August 2002 images. With only two position
measurements, however, the velocities of these stars are subject to
significantly larger errors.

As an additional cross-check of the measured stellar positions, we
compared the time dependent positions of the stars S2 and S8 with the
ones published by \citet{ghez00}, who have used the 10m-class Keck
telescope for a very similar proper motion study (see
Figure~\ref{nttkeck}). In order to take into account the repositioning
of Sgr~A* by \citet{reid03}, we applied an offset of $0.040''$ West
and $0.009''$ North to the Keck data that we calculated from the
differences in the positions of S2 for the 1995 epoch. As can be seen
in Figure~\ref{nttkeck}, the two groups' results are in excellent
agreement.

Our final list of stars with their magnitudes, their positions
relative to Sgr~A* in August 2002, and their proper motion velocities
is given in Table~\ref{list}. All velocities were calculated assuming
a distance of 8~kpc to the Galactic Center \citep{reid93}. An
additional systematic error of the order $\sim$20~kms$^{-1}$ should be
taken into account because of possible systematic errors of
$\sim0.003''$ in the stellar positions (see
section~\ref{orbitsection}). The photometry was calibrated by
selecting suitable reference stars from the \citet{ott03}
list. Magnitudes were measured by aperture photometry on the
Lucy-Richardson deconvolved CONICA/NAOS image from August 2002, with
errors estimated by choosing different aperture sizes.
Figure~\ref{velmap} illustrates the measured stellar proper motions
with velocity vectors superposed on a NACO 2002.4 map.

We identified 6 stars which were subject to significant acceleration
(deviation from a linear trajectory $>3\sigma$) and have marked them
with an asterisk in Table~\ref{list}. The proper motion velocities
given for these stars for the 2002 epoch were derived from  linear fits
to  subsets of the measured positions, i.e. to  time spans that were
short enough to approximate the motions of the stars by linear
trajectories. Therefore, they are approximate estimates and should not
be used for modeling.

As for the naming of some individual sources, S1 through S12 were
named in earlier publications about the central cluster, e.g.
\citet{genzel97}. S12 was mentioned by \citet{genzel97} as a possible
variable source and counterpart of Sgr~A*. \citet{ghez98} detected
this source as well, but excluded the possibility of it being a Sgr~A*
counterpart because of its large proper motion, which was inconsistent
with the expected extremely low proper motion of Sgr~A*. Also, the
repositioning of Sgr~A* by \citet{reid03} moved its location
$\sim50$~mas East of the position used by \citet{genzel00}.  With our
new analysis, based on a much longer time line, we arrive at the
following interpretation: S12 \citep{genzel97} was coincident (at the
level of the SHARP resolution) with a fainter source, S3, in the 1995
epoch.  It passed the pericenter of its orbit at that time, moving at
$>$1000~km/s . The proper motion of S12 was directed towards the
north. In 1998 and 1999 it was located so close to the brighter source
S2 that it could hardly be separated with the resolution of
SHARP/NTT. S12 ``reappeared'' north of S2 in 2000. We dropped S3 from
our list because we could not determine an unambiguous proper motion
for that source.  We show maps for the epochs 1995.5, 1996.4, 1999.5,
and 2000.5 in Figure~\ref{idmaps}. All stars labeled in these maps
(except S3) were subject to significant accelerations as described
below.

\section{Accelerations}

We found that 6 stars in our sample (S1, S2, S8, S12, S13, S14) show
clear signs of nonlinear proper motion, 3 of which are well known
candidates , i.e. S1, S2, and S8 \citep{ghez00, eckart02,
schoedel02}. We measured the accelerations of the five stars S1, S2,
S8, S12, and S13 with parabolic fits to 3 different sections of their
trajectories (in the cases of S2 and S12, sections of the orbits with
small changes in acceleration). The sections were chosen such that the
change of acceleration was negligible for each star. From the 3 fits,
we calculated average accelerations, with corresponding average epochs
and positions. They are listed in Table~\ref{accellist}.  In the case
of S14 the data did not allow such a procedure because its orbit is
seen almost edge on (see Figure~\ref{trajectories}).

In Figure~\ref{accelcones} we show the corresponding acceleration
vectors with their error cones, as determined from the averages of the
three parabolic fits. The radio position of Sgr~A* \citep{reid03} lies
within these error cones for S1, S2, S12, and S8, but not for
S13. However, the latter star is one of the fainter sources in our
sample and has been located very close to S1 between the 1997 and 2000
epochs, which might have deteriorated its measured positions. Its
trajectory will become more precisely determined with future
NACO observations. We therefore excluded S13 from the following
analysis of the location of Sgr~A*.

With the measured projected accelerations of S1, S2, S12, and S8, we
used a maximum likelihood analysis similar to \citet{eckart02} to
determine the position of the dark mass. We used the accelerations and
errors calculated from the three individual parabolic fits to each
star's trajectory. The position of Sgr~A* can be derived from the sum
of the 12 resulting $\chi^{2}$ maps (3 maps for S1, S2, S12, and S8,
respectively). We find from this analyis that the most probable
location of the dark mass is $7^{+12}_{-11}$mas West and
$3^{+21}_{-20}$mas South of the radio position of Sgr~A*. The
confidence limits correspond to $\Delta\chi^{2} = +1$ ($\sim$68\%
probability).

Table~\ref{masslist} lists the enclosed projected masses as derived from
the projected total accelerations. We list values for the inclination
of the orbits of S1, S2, and S8 that were estimated by comparing the
projected mass to an intrinisic black hole mass of
$2.9\pm0.2\times10^{6}$M$_{\odot }$ (see section~\ref{bhmass}).  The
high acceleration of S13 gives a too large enclosed mass and cannot be
used for estimating the inclination. However, it suggests that its
orbit must lie close to the plane of the sky. The enclosed mass
determined for S12 with this method is also too high and does not
allow estimating the inclination of its orbital plane. S12 was close
to bright sources, like S4, S1, and S2, most of the time, so it was in
this case hard to find suitable orbital sections large enough for
fitting well constrained parabolas to them.

\section{Orbits \label{orbitsection}}

The gravitational potential in the central $0.5$~pc is dominated by
that of a point mass \citep{eckart97, genzel00, ghez00, eckart02,
schoedel02}.  If we consider that the mass to light ratio within
$0.55"$ of SgrA*, is comparable to that of the outer cluster ($M/L(2\mu
m)=2\times M_{\odot}/L_{\odot}$), the stellar mass in this radius due to the
stellar cusp will be $\sim5000$~M$_{\odot}$ and the combined
relativistic and Newtonian peri-astron shift of S2 is of the order
$\leq$ 10 minutes of arc per year (Rubilar \& Eckart 2001; Mouawad et
al. 2003). It is therefore reasonable to analyze the motions of the
stars in the Sgr~A* cluster in terms of keplerian orbits.  The four
stars S1, S2, S12, and S14 passed through the pericenter of their
orbits during the time span covered by our observations. In the case of
the three stars S2, S12, and S14 sufficiently large sections of their
orbits were observed in order to allow for a unique keplerian fit. As
for S1, S8, and S13, unique fits were not possible, but we constrained
their orbits, using the estimated inclinations from
Table~\ref{accellist}.

We calculated the orbital parameters by fitting keplerian orbits to
the observed time-dependent positions in the plane of the sky. The
``best'' fit was determined in the least-squares sense. For this
purpose, we built IDL program codes around the IDL MPFIT/MPCURVEFIT
procedures. These procedures use the Levenberg-Marquardt technique to
solve the least-squares problem for a given set of data points and for
a given function. \footnote{The MPFIT program library has been written by Craig
Markwardt and is publicly available at\\ {\em
http://cow.physics.wisc.edu/$\sim$craigm/idl/idl.html}}. If a
significant part of an orbit (of the order $50\%$) has been covered by
observations, it is thus possible to determine its period, semi-major
axis, time of pericenter passage, eccentricity, and the three angles
of orientation, the inclination (except its sign), the angle of the
line of nodes, and the angle from node to pericenter. This is based on
the assumption that the proper motion of Sgr~A* relative to the
surrounding stars can be neglected. With additional spectroscopic
information about line-of-sight velocities (which were not available
for this work) one can also determine the sign of the inclination and
the distance to the GC (Salim \& Gould 1999; Eisenhauer et al. 2003).

We let the initial estimates vary over an appropriately large range in
parameter space in order to ensure finding the global minimum of the
fit. In the cases of S2, S12, and S14, the initial estimates could be
found fairly well from ellipses fitted to their projected
trajectories. In the cases of S1, S8, and S13, where unique fits were
not possible, we calculated their orbital parameters for three
different inclination angles, which were held fixed in the fitting
procedure.  

All results for the orbital parameters are listed in
Table~\ref{orbits}. Only in the case of S2 the position of Sgr~A* was
treated as a free parameter. For the other orbits astrometric errors
of the parameters must be taken into account additionally because of
the error in the radio position of Sgr~A*. They are of the same order
as the fitting errors. The measured time dependent positions of the
stars and their projected orbits as determined by the fits are plotted
in Figure~\ref{trajectories}. \citet{ghez03a} have published
orbital solutions for all the stars described below, which appear to
be in good agreement with our results. As for terminology, they name
the sources S0-1 (S1), S0-2 (S2), S0-4 (S8), S0-16 (S14), S0-19 (S12)
and S0-20 (S13).

\begin{itemize}

\item {\bf S2}: The orbit of S2 has already been reported by
\citet{schoedel02}. Since S2 passed through pericenter early in 2002, the
measurements from that epoch are particularly important for
constraining its orbit. For the analysis of S2, we therefore decided
to include in the case of S2 additional imaging data made with NACO at
the VLT in 2002, i.e. data for the epochs 2002.25, 2002.33, and
2002.58 \citep{schoedel02}.

The NACO AO observations at each epoch provided series of usually
several tens of individual exposures of the GC stellar
cluster. \citet{schoedel02} reported S2 positions for the NACO data
based on the final co-added image for each observing epoch. In this
work, we chose a slightly different approach: S2 is easily identified
in individual exposures without having to apply any
post-processing. Therefore, we measured its position at each NACO
observing epoch by using all available individual exposures. We took
the resulting average position, with the standard deviation of the
individual measurements as an error estimate. These errors are
generally lower than the conservatively estimated errors of
\citet{schoedel02}.

To estimate the systematic error of our SHARP positions, we compared
them with the positions of \citet{schoedel02}, which are taken from
\citet{ott03}, and are based on a different data reduction and
analysis technique. From this comparison we estimated a potential
$\sim3$~mas systematic error on the SHARP positions. We added this
error to our measured positions. The systematic error in position
estimated from this ``cross-calibration'' leads to an additional
systematic error of the order 20~km/s in the proper motion velocities
listed in Table~\ref{list}.

Our analysis of the orbit of S2 is distinct in one other point from
\citet{schoedel02}: We treated the position of Sgr~A* as a free
parameter when fitting the orbit of S2. The resulting orbital
parameters and their formal errors (resulting from an analysis of the
covariance matrix) are listed in column 1 of Table~\ref{orbits}.  We
find a black hole mass of $3.3\pm0.7\times10^{6}$M$_{\odot}$. The
position of the focus of the orbit is $2.0\pm2.4$~mas East and
$2.7\pm4.5$~mas South of the nominal radio position.  This means that
the position of the acceleration center lies well within the $1\sigma$ error
circle of the radio position of Sgr~A*.

Figure~\ref{comporb} compares our data and orbital solution for S2 to
the orbit and positions published by \citet{schoedel02}. We find an
excellent agreement within the errors.  Only the 1992 position seems
to be subject to a larger deviation. We attribute this to the chip
readout pattern in the 1992 data (see section~\ref{sharpdata}). We
also found that weighting of the data is a crucial issue in fitting
keplerian orbits. Therefore, we compare the orbital parameters for
the two cases of normal and equal weighting in
Table~\ref{compweight}. The results agree very well within the errors
and with the values of \citet{ghez03}. However, since the radio
position of Sgr~A* is only known to 10~mas, the offsets of the focus
determined by the two groups are not directly comparable. The
important point is that both groups find that the center of attraction
is offset just a few mas from the radio position of Sgr~A* determined
by \citet{reid03}.

\item {\bf S12}: S12 passed through pericenter in 1995.3. It is a
fairly weak source and has been close to brighter sources, like S2 and
S4, at most of the observing epochs. It passed between S1 and S2 in
1996/97. In 1998 it was almost coincident with S2 and only
``reappeared'' in 1999/2000 north of S2 (see Figure~\ref{idmaps}). For
these reasons, it was not realized earlier that it was a star that had
passed through pericenter. The result of the orbital fit tells us that
it must have been almost coincident with S4 in the 1992
epoch. Photometric measurements support this assumption because S4 is
0.5~mag brighter in the 1992 image, consistent with the sum of the
fluxes of S4 and S12. However, we did not use the 1992 position for
the orbital fit. The black hole mass determined from the orbit of S12
is $3.5\pm2.5\times10^{6}$M$_{\odot}$. This includes the astrometric
error (The focus of the ellipse was not fitted in the case of S12).

\item {\bf S14}: The orbit of this star was first identified by
\citep{ghez03a}, who named this source S0-16. The quality of our data
made the orbit of S14 difficult to determine. We could only measure
reliable positions between the 1998 and 2002 epochs. When fitting the
orbit, we obtained two formal solutions, but disregarded the one that
resulted in an unrealistic black hole mass of
$>10^{7}$M$_{\odot}$.  S14 is a remarkable source because its orbit is
extremely eccentric and is seen almost edge on. The closest point of
approach of this star to the black hole is less than half a light day,
i.e. in principle, it poses even stronger constraints on the central
mass distribution than S2. Unfortunately, we could only determine a
very crude enclosed mass estimate of $6.4\times10^{6}$M$_{\odot}$ with
the error of the same order as this value.

\item {\bf S1, S8, S13}: The orbits of these stars cannot be
  determined with a unique solution yet. However, in order to gain a
  general idea of their orbital parameters we fitted the orbits with
  the inclination angle held fixed at different values, using as an
  orientation the inclination estimates of Table~\ref{masslist}.  The
  results for the three stars are given in Table~\ref{orbits}, with
  superscripts and subscripts being the best fitting values for the
  respective (superscript or subscript) inclination angles.

\end{itemize}

\subsection{Anisotropy}

With our new proper motion data, we looked for possible anisotropy in
the velocity structure of the Sgr~A* stellar cluster, using
$\gamma_{TR} = (v_{T}^2-v_{R}^2)/v^2$ as anisotropy estimator, where
$v$ is the proper motion velocity of a star, with $v_{T}$ and $v_{R}$
its projected tangential and radial components. A value of $+1$
signifies projected tangential motion, $-1$ projected radial motion of
a star. The properties of the anisotropy parameter $\gamma_{TR}$ are
discussed in detail in \citet{genzel00}. They show that an intrinsic
three-dimensional radial/tangential anisotropy will be reflected in
the two-dimensional anisotropy estimator $\gamma_{TR}$.

The errors of the proper motions that are based solely on the Gemini
2000 and the NACO August 2002 data are too large for this analysis and
were therefore not used.  In Figure~\ref{aniso} we have plotted
$\gamma_{TR}$ against the projected distance from Sgr~A* (epoch
2002.7) for the remaining stars from Table~\ref{list}. In
Figure~\ref{anisohist} we show histograms of $\gamma_{TR}$ for stars
within $0.6''$ (dotted lines), $1''$ (straight lines), and $1.2''$ (dashed
lines) of Sgr~A*. Since the stars changed their positions during the
time span covered by the observations, we calculated the anisotropy
parameter for the 2002.5 (lower panel) and for the 1995.5 (upper
panel) epochs. In all cases, the number of stars on radial orbits is
$2-3\sigma$ (assuming Poisson errors) above the number of stars on
projected tangential orbits (cf. Figure~8 of Genzel et al. 2000, who
show histograms of $\gamma_{TR}$ for clusters with varying intrinsic
anisotropy).

From the projected radial velocity dispersions and the ratio of
projected tangential to radial velocity dispersions listed in
Table~\ref{masses} (see section~\ref{bhmass}), we can estimate the
anisotropy of the stellar cluster with the aid of equation~(10) of
\citet{genzel00}:
\begin{equation}
\langle\beta\rangle =
1-\langle\sigma_{t}^{2}\rangle/\langle\sigma_{r}^{2}\rangle =
3(\langle\sigma_{R}^{2}\rangle - \langle\sigma_{T}^{2}\rangle)/(3\langle\sigma_{R}^{2}\rangle - \langle\sigma_{T}^{2}\rangle)
\end{equation}
Averaging the results of the 4 lists in Table~\ref{masses}, we find
$\langle\beta\rangle = 0.5\pm0.2$.  Of course, this value has to be
taken with a certain caution. While the value of the anisotropy agrees
very well for three of the lists, we obtain a value of $\beta\sim0$ in
one out of the four cases. \citet{genzel00} also showed that a
measurement of $\beta$ from a small sample of stars can easily be
skewed towards positive values. However, their simulations showed that
the probability of measuring $\beta\geq0.5$ is as low as $25\%$ for a
sample of $\geq30$ stars out of intrinsically isotropic or
tangentially anisotropic clusters (see their Figure~10).

Additionally, we checked whether any sky-projected overall rotation of
the stars can be detected. Like \citet{genzel03} we used the
normalized angular momentum $J_ {z}/J_ {z}(max) = (xv_y - yv_x)/pv_p$,
where $x$ and $y$ are the offsets from Sgr~A* in R.A. and Decl.,
$v_{x}$ and $v_{y}$ the corresponding velocity components, and $p$ and
$v_{p}$ the projected distance from Sgr~A* and the absolute value of
the proper motion velocity. We show the plot of this parameter against
the projected distance in Figure~\ref{angular}, for the epoch 2002.7
(plotting the parameter for another epoch does not introduce any
significant change).  We find no significant projected overall
rotation of the cluster.

\section{Limits on the emission  of Sgr~A*}

All of the brighter (K$\geq$15.5) sources that can be found in our
maps at some epochs at distances $< 0.1''$ from Sgr~A* are stars with
known proper motions, like S1, S2, or S12. Although fainter sources
(K$\leq$16) can be found within $0.1''$~mas of the location of Sgr~A*
at some epochs as well, they are at least $\sim30$~mas offset from the
exact radio position.  Hence, with the data base used for this work,
we find no clear sign of a stationary or variable
($\geq0.5$~mag) source at the position of Sgr~A*.

Since we have combined the best speckle frames taken during entire
observing runs (which comprised usually several days) for our present
data analysis, this statement does only refer to variability on the
scale of at least several hours to several days. As for shorter term
variability, we did not note any unusual activity at the position of
Sgr~A* when selecting thousands of 0.5~s speckle frames by eye as
described in section~\ref{sharpdata}. However, in order to be clearly
seen in a single SHARP speckle frame, a source must have a K magnitude
of $\sim13.5$ or brighter.

We estimated a simple limit on the average emission of Sgr~A*.  From aperture
photometry with an aperture of $\sim$50~mas and calibration relative
to stars in the cluster we determined a conservative upper limit of
K$>16$ for the emission of Sgr~A* in the epochs before 2000.5 (At that
epoch the bright source S2 moved to within $\sim0.05''$ of Sgr
A*). Using the Galactic Center extinction law of \citet{raab02} (A(K)
= 3.2), we derive an upper limit of $<5$~mJy on the emission from Sgr
A*. This is of the same order as the results of \citet{hornstein02}
and \citet{genzel03}.

\section{Summary and Discussion}

\subsection{The central dark mass \label{bhmass}}

With the proper motions from Table~\ref{list} we calculated the
enclosed mass with the Leonard-Merritt (LM, Leonard \& Merritt 1989)
mass estimator. For this purpose we compiled various lists of stellar
positions and velocities.  We created a long list (LL) with all
sources from Table~\ref{list} and a short list (SL) that excludes the
larger error proper motions determined just from the Gemini/NACO
images. Additionally, we measured the projected velocities of the
accelerated stars at two different epochs in order to take into
account the influence of their varying velocities. We thus obtained 4
lists (SL1, SL2, LL1, LL2). Table~\ref{masses} lists the calculated
values of the enclosed mass, of the radial velocity dispersion, and of
the ratio between the projected radial and tangential velocity
dispersions for all 4 cases.

 \citet{genzel00} showed that the LM mass estimate $\mathrm{M}_{LM}$
differs from the intrinsic mass $\mathrm{M}_{0}$ of the central object
depending on the anisotropy parameter $\beta$ and the power-law slope
of the central stellar cluster. This is because the LM mass estimate
assumes that we have access to the full radial extent of the stellar
cluster, which is not the case for our data set. From Figure~14 of
\citet{genzel00} we estimate that
$0.85<\mathrm{M}_{LM}/\mathrm{M}_{0}<1.0$ for $0<\beta<0.5$ and a
central density slope of $\alpha\approx1.4$ \citep{genzel03}. Taking
into account this range of the corrrection factor, we estimate a
central mass of $3.4\pm0.5\times10^{6}$M$_{\odot}$ from the weighted
average of the LM mass estimates in Table~\ref{masses}. This is fully
consistent with the value for the dark mass obtained from the orbit of
S2.

Figure~\ref{encmass} is a plot of the measured enclosed mass against
distance from Sgr~A*, in close analogy to Figure~17 of
\citet{genzel00} and Figure~3 of \citet{schoedel02}.  The main
differences to \citet{schoedel02} are: (1) The error of the mass
estimate from the orbit of S2 has been reduced by taking the position
of the orbital focus explicitly into account. (2) The innermost LM
mass estimate was based on the \citet{ott03} data in
\citet{schoedel02}. It has been replaced by the LM mass estimate from
the present work, which is based on more  data  in the
region within $\sim1''$ of Sgr~A*. (3) The LM mass estimates in
Figure~3 of \citet{schoedel02} and Figure~17 of \citet{genzel00} were
corrected downward by 5-10\% because they assumed a power-law slope of
$\alpha=1.8$ for the inner stellar cluster. Now we use a power-law
slope of $\alpha \approx 1.4$ \citep{genzel03}. This means that the LM
mass estimates were previously  underestimated by $\sim$10\% .

 Fitting a model composed of a point mass plus the visible stellar
cluster with a core radius of 0.34~pc and a power-law slope of
$\alpha=1.8$ to the data gives a value of
$2.9\pm0.2\times10^{6}$M$_{\odot}$ for the central dark mass. This
agrees within the errors with the LM mass estimate of the innermost
stars and with the masses calculated from the orbital parameters of S2
and S12. The values for the point mass given by \citet{chakrasaha01}
and \citet{ghez98} are systematically lower than the other
estimates. If we disregard them in our analysis, we obtain a central
mass of $3.1\pm0.2\times10^{6}$M$_{\odot}$. In order to provide an
overview over the different mass estimates, we have compiled some of
them in Table~\ref{bestmasses}. From their analysis of the orbit of
S2, \citet{ghez03} measured the currently highest value of
$4.1\pm0.6\times10^{6}$M$_{\odot}$. It is the most extreme value, but
could for example be realized if the central cluster were isotropic
and its power-law index $\alpha\approx1.3$ (see Figure~14 of
\citet{genzel00}).

The mass of Sgr~A* can also be constrained via its proper motion
measured by radio astronomical observations. \citet{reid03a} find an
upper limit of 8~kms$^{-1}$ for the intrinsic proper motion of
Sgr~A*. Following their argument, we can apply equipartition of
momentum to the case that the black hole is perturbed by close
passages of stars. For S2, with a mass of $\approx15$~M$_{\odot}$ and
a velocity of $\approx7000$kms$^{-1}$ at pericenter, we find a lower
limit of $13\times10^{3}$~M$_{\odot}$. \citet{dorband03} model the
Brownian motion that a supermassive black hole would have embedded in
the stellar cluster at the center of the Milky Way. From their
equation (73) we calculate a minimum mass of Sgr~A* of
$7.5\times10^{3}$~M$_{\odot}$ with the above proper motion
constraint. Carrying the argument a bit further, we can interprete the
upper limits on the size of Sgr~A* at millimeter wavelengths as its
half mass diameter.  Assuming a half mass radius of $r_{h}\leq0.1$mas
\citep{doeleman01, rogers94, meliafalcke01} we obtain with the above
mass estimates
\begin{equation}
\rho_{SgrA*} \geq
\frac{7.5\times10^{3}\mathrm{M}_{\odot}/2}{(4\pi/3)r_{h}^{3}} >
3\times10^{19} \mathrm{M}_{\odot}\mathrm{pc}^{-3}
\end{equation}

Figure~\ref{darkmass} is a summary of the lower limits on the density
of the central dark mass obtained by different analyses. Considering
our data, S2 places currently the tightest constraints on the central
mass distribution. Although S14 approaches the dynamical center even
closer, our data allow not to constrain its orbit with the necessary
precision. The upper limit on the proper motion of Sgr~A* relative to
the surrounding star cluster gives the overall highest mass density
constraint.  If we assume a Plummer model cluster of dark
astrophysical objects instead of a central point mass, its central
density must exceed $2.2\times10^{17}$M$_{\odot}$pc$^{-3}$ as shown in
Figure~\ref{encmass}. The lifetime of such a cluster is limited to
less $<10^{5}$~years \citep{maoz98}. A single fermion ball model that
explains all the observed central dark mass concentrations at the
centers of galaxies can also be excluded
\citep{schoedel02}. Considering that a boson star \citep{torres00}
should eventually collapse to a black hole through accretion of the
abundant gas and dust in the GC, we conclude that current models and
measurements compellingly suggest the presence of a supermassive black
hole at the center of the Milky Way. As for a tight ($<10$ light hours
separation) binary black hole with similar masses of its two
components, it would coalesce by gravitational radiation within a few
hundred years (B.F. Schutz 2003, private communication).

\subsection{Anisotropy of the central cluster?}

We used the measured projected stellar velocities and the anisotropy
estimator $\gamma_{TR} = (v_{T}^2-v_{R}^2)/v^2$ to look for signs of
anisotropy of the stellar cluster in the immediate vicinity of
Sgr~A*. We found a predominance of $\gamma_{TR}=-1$ on the $2-3\sigma$
level. This is of moderate statistical significance because of the
small number of stars, but it is a stable result as to variation in
epoch and in the size of the examined region around Sgr A*. The
possible radial anisotropy appears to become more significant with
shorter projected distances from the central black hole. No sign of a
significant projected overall rotation is detected.

\citet{genzel00} based their claim of radial anisotropy in the inner
$0.8''$ near Sgr~A* on the absence of stars with $\gamma_{TR} =
+1$. Although we now find some stars with $\gamma_{TR} = +1$ in our
new proper motion data, this does not contradict a possible radial
anisotropy. As \citet{genzel00} show, the distribution of
$\gamma_{TR}$ is always bimodal, so even in the case of radial
anisotropy we expect some stars with $\gamma_{TR}=+1$. Hence, we feel
that our analysis provides sufficient evidence for radial anisotropy
of the sources brighter than $\mathrm{K}\sim16$. Whether this holds for the
fainter sources as well, must be checked with the upcoming deep AO
observations (see section~\ref{outlook}). We suggest the following
tests of anisotropy:

\begin{itemize}

\item The most obvious approach to the problem is measuring the proper
  motions of a greater number of stars. Observations with NACO at the
  VLT will provide such data on the fainter stars in the Sgr~A*
  cluster within the next few years. We expect that the number of
  measured proper motions within $1''$ of Sgr~A* will be augmented by
  a factor of 2 to 3.

\item An alternative method to test the isotropy/anisotropy of the
 central stellar cluster is the eccentricity of stellar orbits: In a
 spherical system of test particles orbiting a point mass we expect
 that the number of particles with eccentricities in the range $(e,
 e+de)$ is proportional to $ede$ (Binney and Tremaine, p. 282, problem
 4-22). We show the corresponding cumulative distribution function in
 Figure~\ref{epdf}. It is skewed towards high eccentricities: 75\% of
 the stars have $e>0.5$, 19\% have $e>0.9$ and 10\% have $e>0.95$. In
 the case of radial anisotropy, we expect even more stars on highly
 eccentric orbits because stars with specific energies $E$ in an
 interval $(E, E+dE)$ would have a lower average angular momentum than
 in the isotropic case.

 We included the eccentricities of the 6 orbits examined in
 section~\ref{orbitsection} into Figure~\ref{epdf}. All points lie to
 the right and below the theoretical curve. This might be a hint that
 there are indeed more highly eccentric stellar orbits than in the
 isotropic case \emph{if} there is no bias to detect preferentially
 orbits with high eccentricities.  When examining such a possible bias
 in section~\ref{bias}, we find that eccentricities $e>0.9$ could be
 subjected to such a bias. However, the number of measured
 eccentricities is still too low for a conclusive test, but this
 method might provide a useful for future observations.

\item In principle, the anisotropy of the central stellar cluster can
  also be measured by comparing the mass of the central black hole as
  determined from LM mass estimates to the mass determined from
  keplerian orbital fits.

  \citet{genzel00} showed that the ratio between the true mass of the
  black hole at the center of the cluster and its LM mass estimator
  depends on both the power-law exponent $\alpha$ of the density of the
  central cluster and on its anisotropy parameter $\beta$ (see their
  Figure~14). If one assumes the validity of keplerian orbits near
  Sgr~A*, it is thus possible to determine $\beta$ by comparing
  sufficiently precise mass measurements from individual orbits and LM
  mass estimators from analyses of sufficiently large proper motion
  samples. At the moment, the errors on the involved parameters still
  allow for a range of $0\leq\beta\leq0.5$. Future
  measurements, however, should set tighter limits.
\end{itemize}

With densities of the order $10^{7}$ to $10^{8}$~M$_{\odot}$pc$^{-3}$
in the central arcsecond \citep{genzel03}, two-body interactions
should be comparably frequent and the relaxation time is less than
$10^8$~yr. In this case we would expect to observe an isotropic
velocity field. There is a possibility that only the brighter,
possibly young stellar component is unrelaxed and is characterised by
radial anisotropy.  It is clear that we need a larger and deeper
proper motion data base before we can definitely exclude isotropy of
the overall cluster. But should the radial anisotropy indeed be proven
to be true, theoretical and modeling efforts will be needed to
understand this property of the Sgr~A* stellar cluster.

As a bottom line, we can say that the current data definitely exclude
tangential anisotropy. Significant tangential anisotropy would be
expected in systems with binary black holes \citep{gebhardt02}. Stars
on highly eccentric orbits would be ejected or destroyed
preferentially in such systems.

\subsection{Biased detection of orbits? \label{bias}}

In the previous section we saw that we could learn about the anisotropy
characteristics of the dense stellar cluster surrounding Sgr~A* from
the measured eccentricities of the observed orbits. However, at this
point the question arises whether there is a bias to detect
preferentially  circular or highly eccentrical orbits. 

When observing stellar proper motions/accelerations there is always a
given lower limit for the detectable acceleration imposed by the
limitations of the telescope, instrument, and generally by the
observing technique (e.g. time sampling).  Consider two stars on
keplerian orbits of equal energy (equal semi-major axis) around a
black hole. The more eccentric of the two will spend less of its time
near the black hole where its acceleration is detectable, and its
eccentricity will therefore have a smaller probability of being
measured. This effect biases the detected eccentricities towards
smaller values. On the other hand, as long as $\alpha < 2$ the number
of stars per interval $da$ increases with $a$ (Appendix eq 4). These
more numerous high-$a$ stars can have large detectable accelerations
only if their eccentricity is high enough to take them close to the
black hole. This effect biases the detected eccentricities toward
higher values. Thus, there are two competing trends that may introduce
a bias to the measured distribution.

In order to obtain a rough quantitative estimate of the possible bias,
we modeled families of keplerian orbits with fixed eccentricities and
specific energies. We let their inclination angle $i$ and the angle
from node to pericenter $\omega$ (i.e. the rotation of the orbit in
the plane of the orbit) vary on grids. $\omega$ was varied between $0$
and $2\pi$ in equal steps of $\pi/6$, while $sin(i)$ was varied in
equal steps of $0.1$ between $-0.9$ and $0.9$. For each of the
individual orbits we calculated the fraction of the orbital period
that the acceleration as seen projected onto the plane of the sky
stayed above a given threshold value. We interpreted this fraction as
the probability of detecting the given orbit. We chose as a detection
threshold a minimum acceleration of 1~mas~yr$^{-2}$, a value slightly
lower than the lowest detected acceleration in our data sample (star
S8). By averaging the probabilities over all angles $i$ and $\omega$,
we obtained the overall detection probabilities for orbits with a
given semimajor axis $a$ and a given eccentricity $e$. We assumed an
isotropic cluster and that the probability distribution function for
$e$ was independent of $a$. We assumed a time sampling of two
observations per year.

That leaves creating an adequate distribution of semimajor axes. We
chose a minimum semimajor axis corresponding to an orbital period of
$\sim$5~yr, which we consider a realistic value for our data
sample. The upper limit for the distribution of semimajor axes is
chosen to be 1~pc, roughly the distance over which the black hole
dominates over the mass of the stellar cluster.  The energy
distribution function near a massive black hole has the form
\begin{equation}
f(\epsilon) =  A\epsilon^{p} \quad
n_{*}\propto r^{-3/2-p} \quad p = \frac{M}{4M_{2}}
\end{equation}
if the stellar population consists of a spectrum of masses
$M_{1}<M<M_{2}$, where $\epsilon=-v^{2}/2+GM_{BH}/r$ is the specific
energy of the star and $f_{M}\equiv0$ for $\epsilon<0$ (Bahcall \&
Wolf 1977; see also Appendix~A). $n_{*}$ is the stellar number
density. The distribution function of semimajor axes $n(a)$ for a
given cusp model is then $n(a)\propto a^{1/2-p}$ (see appendix~A). We
created distributions of semimajor axes (1000 values of $a$ sampled
according to $n(a)$ between its minimum and maximum values)
corresponding to isotropic power-law stellar cusps with exponents
$\alpha$ of $-1.0$, $-1.5$, and $-2.0$.  In the case of the Milky Way
black hole the stellar density is approximately $n_{*}\propto
r^{-3/2}$ \citep{genzel03}.

Figure~\ref{biasplot} shows the probabilities of detecting orbits with
a given $e$ and $a$ for the case of a $n\propto -1.5$ cusp. We only
plotted the curves for orbits with $100\quad\mathrm{mpc}\gtrsim a$
because the plot would get extremely crowded at large semimajor axes
and because the probabilities for large $a$ are extremely low.  For
small $a$, the detection probability is large for all $e$, while at
intermediate $a$, there is a bias towards less eccentric orbits. This
bias turns over at large semimajor axes, when stellar accelerations
can only be observed near the pericenter of highly eccentric
orbits. However, for orbits with large $a$, detection probabilities
are extremely low.

By averaging over all semimajor axes we obtained the overall detection
bias of orbits with a given eccentricity in a given isotropic cusp
model. The results are plotted in Figure~\ref{ebias}. The exact values
of the probabilities depend on the modelling parameters (e.g. the
number of steps when creating the distribution of semimajor
axes). However, we are only concerned with the relative probabilities.
In the model with $\alpha=-2.0$ the stars are most concentrated toward
the center and therefore the overall probabilities of detecting orbits
are highest. In this case, there is no significant overall detection
bias for any eccentricity. At the other extreme, in case of
$\alpha=-1.0$, there is a bias factor of $>2$ for highly eccentric
orbits compared to circular orbits. This corresponds to our
expectation that the relative weight of more circular orbits should
decrease with a decreasing concentration of the cluster. The case of
$\alpha=-1.5$, which is closest to the case of the GC stellar cusp,
shows some bias toward high eccentricities. However, this bias is
modest (factor $1.5-2$). Basically, it only affects significantly
extreme eccentricities of $e>0.9$.

With our modeling we wanted to provide a general idea of the possible
biases involved in detecting stellar orbits around Sgr~A*.  As for the
eccentricities of the stellar orbits found in the present work, out of
the six stars two stars, i.e. S1 and S8, have semimajor axes in the
regime where circular orbits are only detected with low
probabilities. As for the orbits of the remaining four stars, they
have semimajor axes that would favour the detection of more circular
orbits. So there might be a bias on our sample, but the small number
of orbits is not sufficient for a conclusive test.

\subsection{No detection of a NIR counterpart of Sgr~A*}

We measured a conservative limit of 5~mJy on the dereddened emission
of Sgr~A* at $2.2\mu$m, in agreement with similar measurements
\citep{hornstein02, genzel03}. We found no evidence for a varying
and/or stationary source at the position of Sgr~A* in our
maps. However, this addresses only time scales of at least several
hours to days. Also, no variability at the position of Sgr~A* was
found in the course of the ``by eye'' inspection of thousands of
speckle frames for every observing epoch. However, this method, where
we could have picked up variability on time scales of seconds up to
$\sim10$ minutes, is limited by the short 0.5~s single frame
integration times to K fluxes of $\sim13.5$ or brighter. As for
\citet{hornstein02}, they sampled their data into 3~hour periods and
found no significant variability. Combined with our present work, that
leaves open time scales from seconds to several tens of minutes, at K
magnitudes fainter than $\sim13.5$.

In the light of the results of \citet{genzel03}, who found the
predicted stellar cusp around Sgr~A*, it is worthwhile to point out
that the probability of detecting a $K\leq17$ star at a projected
position of $<0.1''$ from Sgr~A* is close to one. With a number
density of $>$50~stars arcsec$^{-2}$ at distances $<0.5''$ (see
Figure~7 of \citet{genzel03}), we find that the average distance
between stars in the central region is $<0.1''$, when only taking into
account stars brighter than $\mathrm{K}\sim17$.

In fact, Figure~7 of \citet{genzel03} suggests that even with an
8m-class telescope, the confusion limit might be reached within
$0.1''$ of Sgr~A* in deep (K$\sim$19) NIR imaging
observations. Additionally, if a NIR source at the position of Sgr~A*
could be detected, one would still have to measure its proper motion
and photometric properties over several epochs and/or perform
spectroscopy or polarimetry in order to clarify its nature. Hence,
detecting Sgr~A* in the NIR seems to be a difficult task.

However, current models of the accretion flow predict that the
infrared emission comes from the tail of the non-thermal radio/submm
emission, and not from a hot, thermal accretion disk
\citep{markoff01,liumelia02}.  Therefore, attempts to detect Sgr A* at
longer wavelengths, where the emission from hot stars is also
significantly lower, seem more promising and will provide tighter
constraints on the emission models.

\subsection{Outlook \label{outlook}}

The stellar cusp in the immediate vicinity of Sgr~A* provides the
opportunity to study the interactions between its components and the
central supermassive black hole on a level of detail that will never
be achievable in other galaxies. It poses a number of intriguing
astrophysical problems, such as its possible radial anisotropy and the
theoretically unexpected presence of young, early-type stars
\citep{ghez03, genzel03}. Observations in the next years will enable
us to examine the Sgr~A* cluster in unprecedented detail and hopefully
solve some of its enigmas.

In the H-band, NACO at the VLT provides images with a resolution
$\sim$3 times higher than SHARP/NTT, reaching of the order 3
magnitudes deeper. In Figure~\ref{gcnaco}, we show a Lucy-Richardson
deconvolved NACO H-band image (reconvolved with a gaussian beam of
FWHM $\sim$40~mas) of 1500~s total integration time. We have counted
``by eye'' $\sim$90 stellar sources with magnitudes H$\leq$18 within
$1''$ of Sgr~A*. There are of the order 20 stars within $0.5''$ of
Sgr~A* with poorly known or unknown proper motions. All of these
sources are candidates for measuring accelerations and constraining
the properties of stellar orbits near the supermassive black hole.  We
will thus be able to build a proper motion data base with much
improved statistics.  It will then be possible to settle the question
of the isotropy/anisotropy of the central stellar cluster under
consideration of the faint component.

We expect also to determine stellar orbits with a much higher
precision and to detect several faint stars with orbital periods of
the order 10-20 years. This will allow to locate the center of
acceleration to wihtin $<$1~mas and to measure the mass of the black
hole precisely and without the assumptions involved in statistical
methods. Measuring the radial velocity of the orbiting stars allows a
precise determination of the distance to the Galactic Ceter
\citep{salimgould99,ghez03,eisenhauer03}. We expect that stars will be
found orbiting Sgr A* with velocities $\geq0.1\times c$, approaching
the relativistic regime.

Spectroscopy of sources in the dense stellar cluster with AO slit
spectroscopy \citep{ghez03} and AO integral-field units
(SPIFFI/SINFONI at the VLT) will help us understand the nature of the
stars near the black hole and their three-dimensional velocity and
space distribution. Finally, imaging and polarimetric observations at
longer wavelengths will help constrain the parameters of the accretion
flow onto Sgr~A*.

\acknowledgments

We like to thank the ESO NTT team for their help and support during
ten years of observations with the SHARP guest instrument.  

We thank the NAOS and CONICA team members for their hard work, as well
as the staff of Paranal and the Garching Data Management Division for
their support during the commissioning and science verification of
NACO.

We thank Andrea Ghez of UCLA for helpful discussions and exchange of
opinions on the stellar orbits, especially for understanding the orbit
of S14/S0-16.

Rainer Sch\"odel thanks William D. Vacca for his great help on fitting
techniques.

TA is supported by GIF grant 2044/01, Minerva grant 8484,
and a New Faculty grant by Sir H. Djangoly, CBE, of London, UK.

Based in part on observations obtained at the Gemini Observatory, which is
operated by the Association of Universities for Research in Astronomy,
Inc., under a cooperative agreement with the NSF on behalf of the
Gemini partnership: the National Science Foundation (United States),
the Particle Physics and Astronomy Research Council (United Kingdom),
the National Research Council (Canada), CONICYT (Chile), the
Australian Research Council (Australia), CNPq (Brazil) and CONICET
(Argentina).

\appendix
\section{The spherically symmetric power-law cusp}

We present here for convenience a self contained summary of some properties
of the distribution function (DF) of a spherically symmetric system,
and apply them to a power-law cusp and keplerian orbits.

\subsection{Isotropic velocity field}

The DF $f(\varepsilon)$ of a system with an isotropic velocity field
is a function of the specific energy only,
$\varepsilon\equiv-v^{2}/2+\psi(r)$, where $\psi\equiv-\phi$ and
$\phi$ is the gravitational potential ($f$ has units of
$x^{-3}v^{-3}$; $f(\varepsilon)\!=0$ for $\varepsilon\le0$).  The
space density distribution, $n(r)\equiv\int
f(\varepsilon)\mathrm{d^{3}}v$ (in units of $x^{-3}$), is
\begin{equation}
n(r)=4\pi\int_{0}^{\sqrt{2\psi(r)}}f(\varepsilon)v^{2}\mathrm{d}v\mathrm{=4\pi\int_{0}^{\psi(r)}}f\mathrm{(\varepsilon)\sqrt{2(\psi-\varepsilon)}}d\varepsilon\mathrm{\,,}\label{eq:nr}\end{equation}
where the last step is made with the variable transformation
$\mathrm{d}\varepsilon=-v\mathrm{d}v$, and
$v=\sqrt{2(\psi-\varepsilon)}$.

For a power-law DF, $f(\varepsilon)=A\varepsilon^{p}$, \begin{equation}
n(r)=(2\pi)^{3/2}\frac{\Gamma(1+p)}{\Gamma(5/2+p)}A\psi(r)^{3/2+p}\,.\label{eq:nrpl}\end{equation}
Very near the black hole the orbits are keplerian to a good approximation,
$\psi\!=\! Gm/r$, and therefore $n(r)\propto r^{-3/2-p}$. 

The distribution of specific energy is $n(\varepsilon)\equiv\int f(\varepsilon^{\prime})\delta(\varepsilon^{\prime}-\varepsilon)\mathrm{d^{3}}x\mathrm{d}^{3}v$
(in units of $v^{-2}$), where $\delta$ is the Dirac delta function
(in units of the inverse of its argument). Using the property $\delta(h(x))=\delta(x-x_{0})/\left|\mathrm{d}h/\mathrm{d}x\right|$,
where $h(x_{0})=0$, and performing first the integration over velocities,
one obtains for keplerian orbits $(a=Gm/2\varepsilon)$\begin{eqnarray}
n(\varepsilon) & = & (4\pi)^{2}f(\varepsilon)\sqrt{Gm}\int_{0}^{2a}\sqrt{2/r-1/a}r^{2}\mathrm{d}r\nonumber \\
 & = & \sqrt{2}\pi^{3}(Gm)^{3}f(\varepsilon)\varepsilon^{-5/2}\nonumber \\
 & = & 2^{3-p}\pi^{3}(Gm)^{1/2+p}Aa^{5/2-p}\,.\end{eqnarray}
The distribution of semi-major axes, $n(a)$ (in units of $x^{-1})$
is therefore \begin{equation}
n(a)=\left|d\varepsilon/da\right|n(\varepsilon)=2^{2-p}\pi^{3}(Gm)^{3/2+p}Aa^{1/2-p}\,.\end{equation}
The distribution of eccentricities in a spherically symmetric distribution
of keplerian orbits with an isotropic velocity field is derived below
(Eq. \ref{eq:ne}).

\subsection{Anisotropic velocity field}

In the general spherically symmetric case, the DF $f(\varepsilon,L)$
depends on both the specific energy $\varepsilon=-v_{r}^{2}/2-L^{2}/2r^{2}+\Psi$
and the magnitude of the specific angular momentum, $L^{2}=(rv_{t})^{2}$,
where $v_{r}$ and $v_{t}$ are the radial and transverse components
of the velocity relative to the radius vector.

In cylindrical coordinates $\mathrm{d}^{3}v=2\pi v_{t}\,\mathrm{d}v_{t}\,\mathrm{d}v_{r}$,
and so substituting $v_{t}\,\mathrm{d}v_{t}=L\mathrm{d}L/r^{2}$,
$\mathrm{d}v_{r}=-\mathrm{d}\varepsilon/v_{r}$ with an extra factor
of 2 to account for both contributions from $\pm v_{r}$ to $\mathrm{d}\varepsilon$,
one obtains\begin{equation}
\mathrm{d}^{3}v=4\pi L\,\mathrm{d}L\,\mathrm{d}\varepsilon\left/r^{2}\left|v_{r}(\varepsilon,L,r)\right|\right.\,.\end{equation}

The distribution of specific energy and angular momentum, $n(\varepsilon,L)$
(in units of $x^{-1}v^{-3}$)is obtained by integrating $f(\varepsilon,L)\,\mathrm{4}\pi r^{2}\mathrm{d}r\,\mathrm{d}^{3}v/\mathrm{d\varepsilon d}L$
over the range $(r_{-},r_{+})$ that is accessible with $\varepsilon$
and $L$,\begin{eqnarray}
n(\varepsilon,L) & = & 16\pi^{2}Lf(\varepsilon,L)\int_{r_{-}}^{r_{+}}\frac{\mathrm{d}r}{\left|v_{r}(\varepsilon,L,r)\right|}\nonumber \\
 & = & 8\pi^{2}Lf(\varepsilon,L)T_{r}(\varepsilon,L)\,,\label{eq:NeL}\end{eqnarray}
where $T_{r}$ is the radial period and the integral expresses the
contribution from each $\mathrm{d}r$ segment of the orbit, weighted
by the time spent there.

For keplerian orbits $\varepsilon\!=\! GM/2a$, $L^{2}\!=\! GMa(1-e^{2})$
and $T_{r}\!=\! P(\varepsilon)\!=\!\pi GM/\sqrt{2\varepsilon^{3}}$,
and so $2L\,\mathrm{d}L=-(GM)^{2}/(2\varepsilon)e\,\mathrm{d}e$.
It then follows from Eq. (\ref{eq:NeL}) that the distribution of
specific energy and eccentricity (in units of $v^{-2}$) is (cf Cohn
\& Kulsrud 1978)

\begin{equation}
n(\varepsilon,e)=\left[2\sqrt{2}\pi^{3}(GM)^{3}f(\varepsilon,L)\varepsilon^{-5/2}\right]e\,,\end{equation}
It then follows that for keplerian orbits with \emph{isotropic} velocities
the distribution of eccentricities $n(e)$ (dimensionless) is (cf
Binney \& Tremaine 1987)

\begin{equation}
n(e)=\left[2\sqrt{2}\pi^{3}(GM)^{3}\int f(\varepsilon)\varepsilon^{-5/2}\,\mathrm{d}\varepsilon\right]e\propto e\,.\label{eq:ne}\end{equation}

\clearpage

\begin{figure}
\plotone{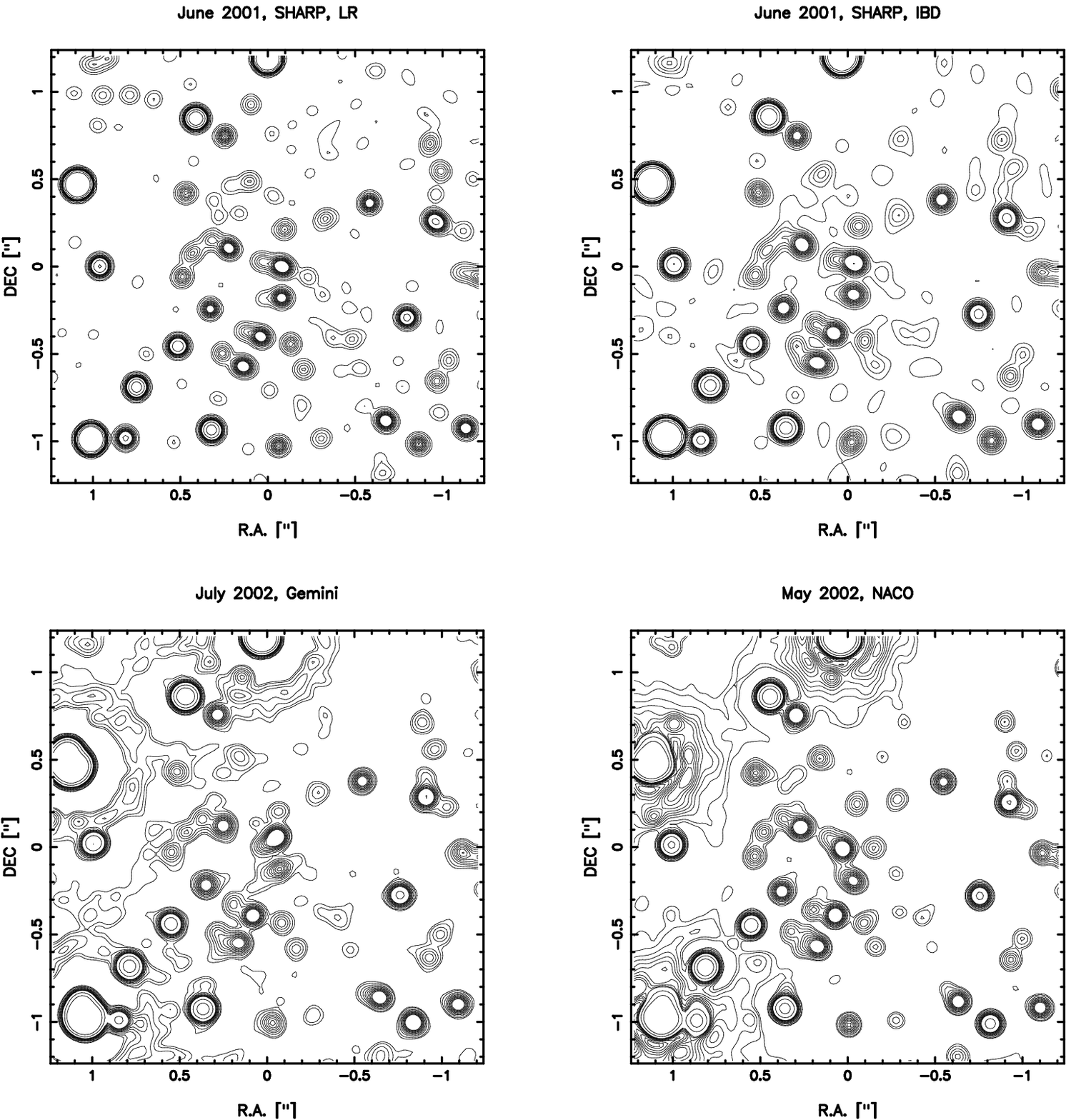}
\caption{A comparison of high-resolution maps of the central $\sim
  1.2''\times1.2''$ of the GC stellar cluster. The restoration beam
  FWHM was approximately 100 mas in all the images. Upper left:
  SHARP/NTT 2001, deconvolved with LR; upper right: SHARP/NTT 2001,
  deconvolved with IBD; lower left: Gemini 2000 (LR); lower right:
  NACO 2002 (LR). The bright sources at the left and upper edges of
  the field were saturated in the latter two data sets and show prominent
  deconvolution artifacts in form of rings. Contours at 10, 20, \ldots
  90, 100, 200, 400\% of the peak flux of S2, the bright source near
  the very center of the images.
\label{comparison}}
\end{figure}

\begin{figure}
\plotone{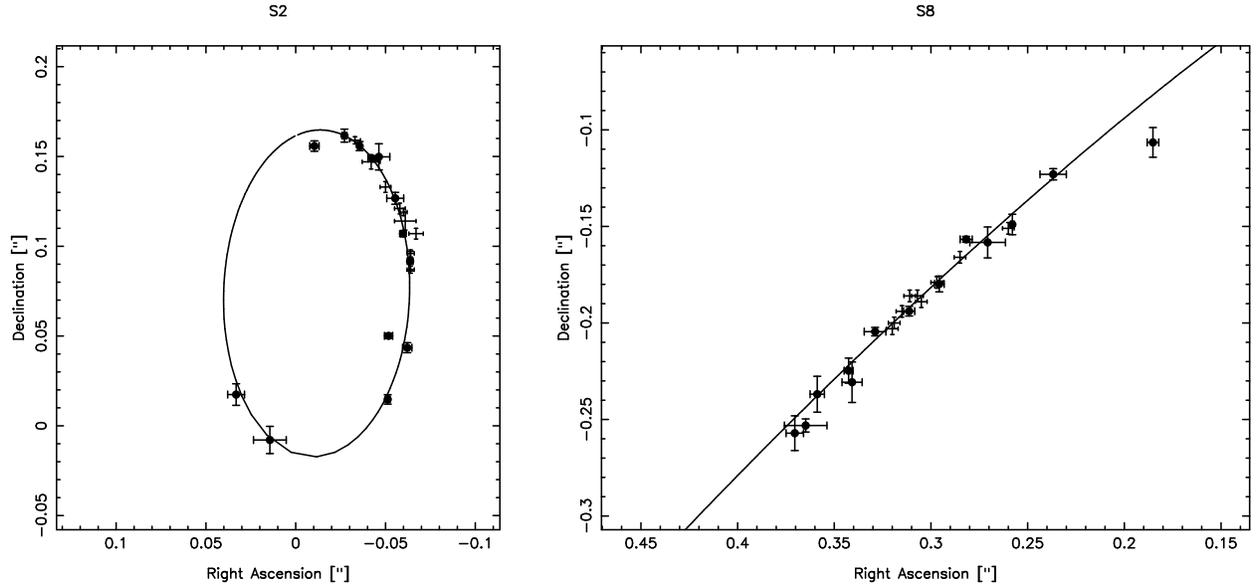}
\caption{A comparison between positions of the stars S2 and S8 as
  measured by our group and by the Ghez et al. (2000) group. SHARP/NTT
  positions are marked by filled black circles with error bars, Keck
  positions just by their error bars. An offset of 40~mas W and 9~mas
  N (derived form the difference in position of S2 for the 1995 epoch)
  was applied to the Keck data in order to take into account the
  astrometric offset between the data sets. Straight lines mark the
  orbits fitted to the position of S2 and S8 (see
  section~\ref{orbitsection}). \label{nttkeck}}
\end{figure}

\begin{figure}
\plotone{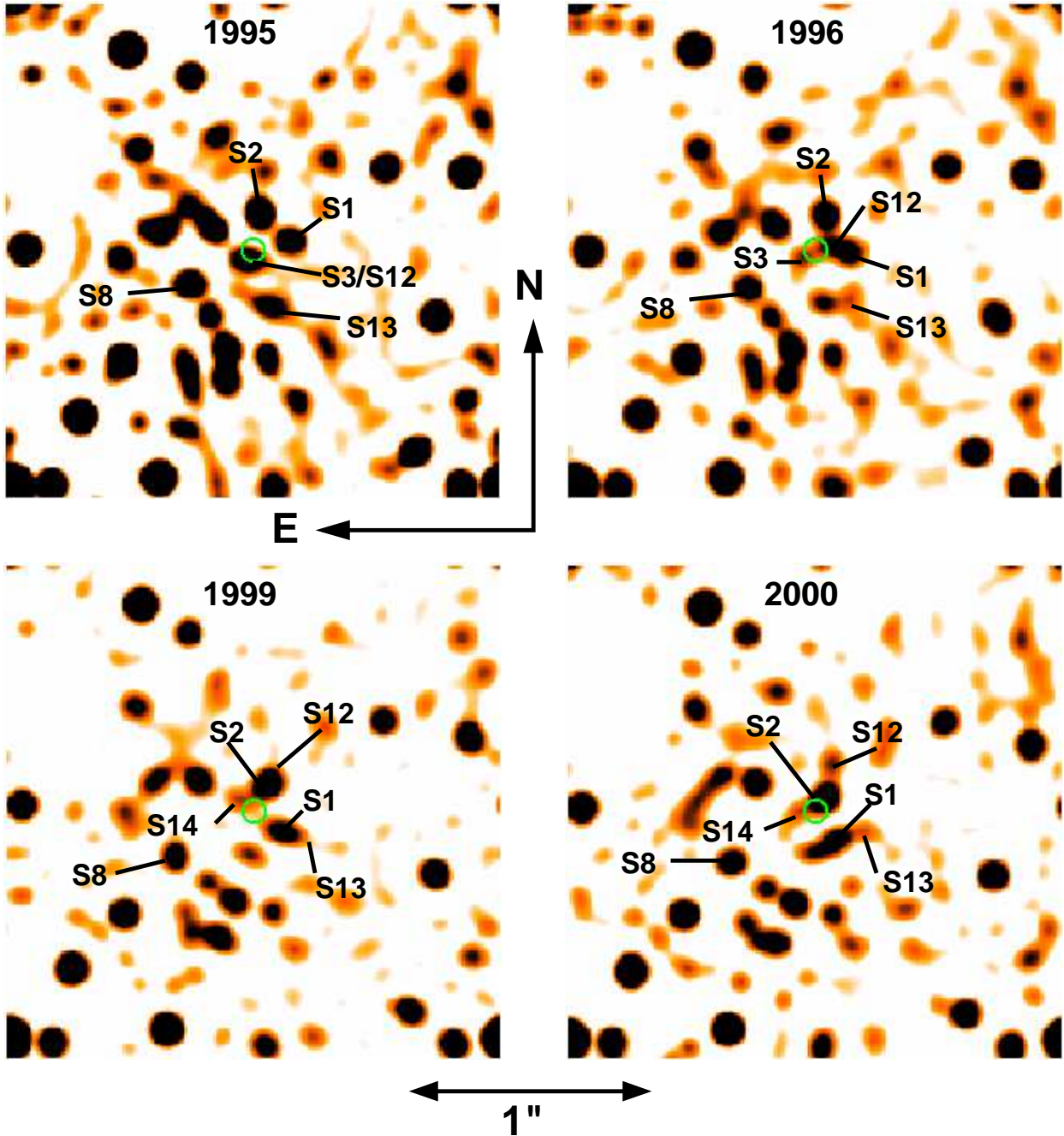}
\caption{Identification of the sources S1, S2, S3, S8, S12, S13, and
  S14 for the epochs 1995.5, 1996.4, 1999.5, and 2000.5. The circle of
  $\sim$50~mas radius marks the position of Sgr~A*. \label{idmaps}}
\end{figure}

\begin{figure}
\includegraphics[height=\textwidth,angle=270]{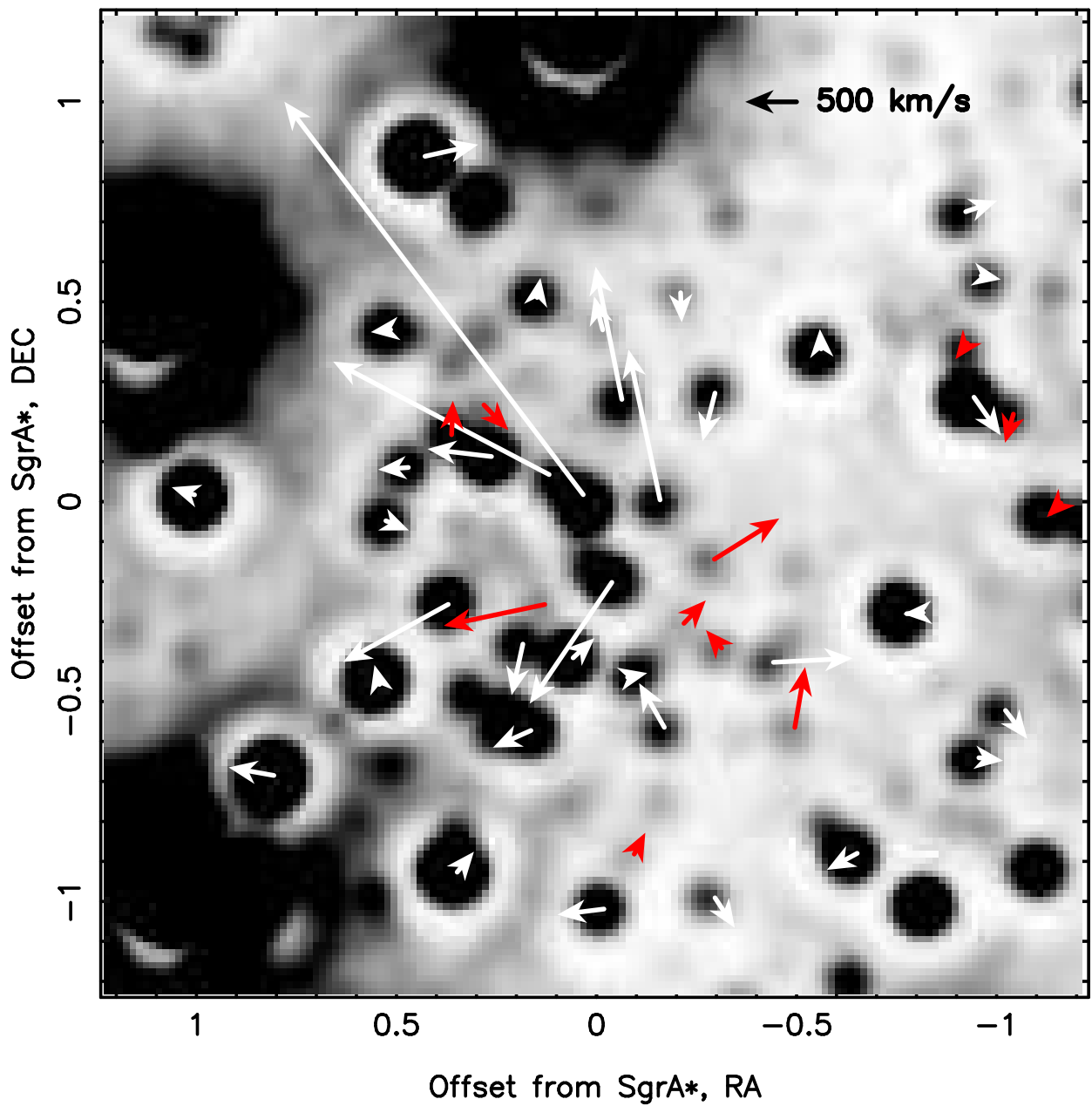}
\caption{Stellar velocities within $1.2''$ of Sgr~A*, superposed on a
  NACO 2002.4 image. Proper motions of the accelerated sources S1, S2,
  S8, S12, S13 and S14 are approximate estimates for this epoch. White
  arrows: Stars with proper motions based on the entire data set. Grey
  arrows: Stars with proper motions determined from the Gemini 2000
  and the NACO August 2002 images only. \label{velmap}}
\end{figure}

\begin{figure}
\includegraphics[height=\textwidth,angle=270]{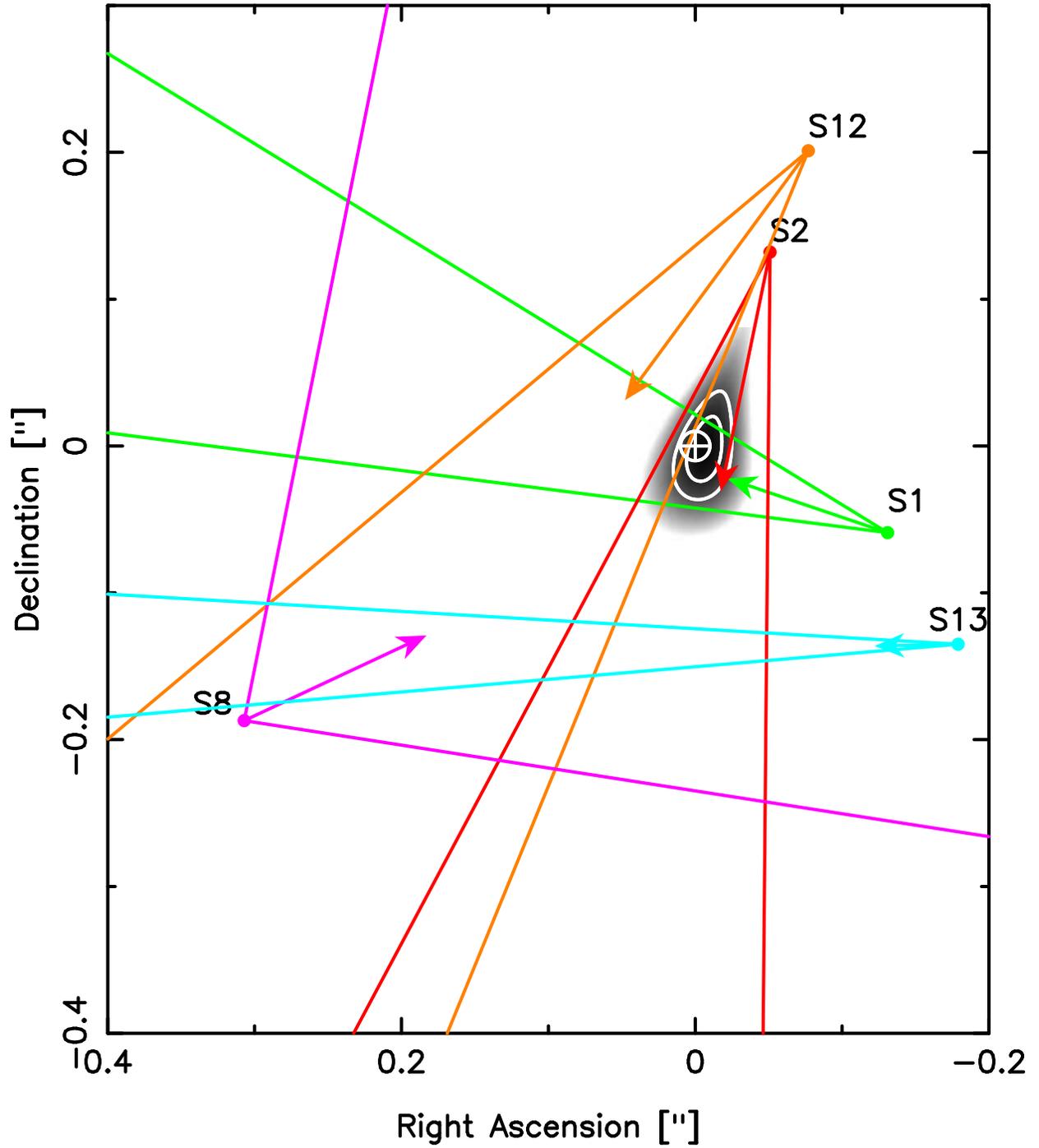}
\caption{Projected average acceleration vectors with their $2\sigma$
error cones for the stars S1, S2, S8, S12, and S13. A $\chi^{2}$ map
of the position of the center of acceleration is overlaid in grey
shades. White contours designate the 1 and 2$\sigma$ confidence
levels. \label{accelcones}}
\end{figure}

\begin{figure}
\includegraphics[width=\textwidth,angle=270]{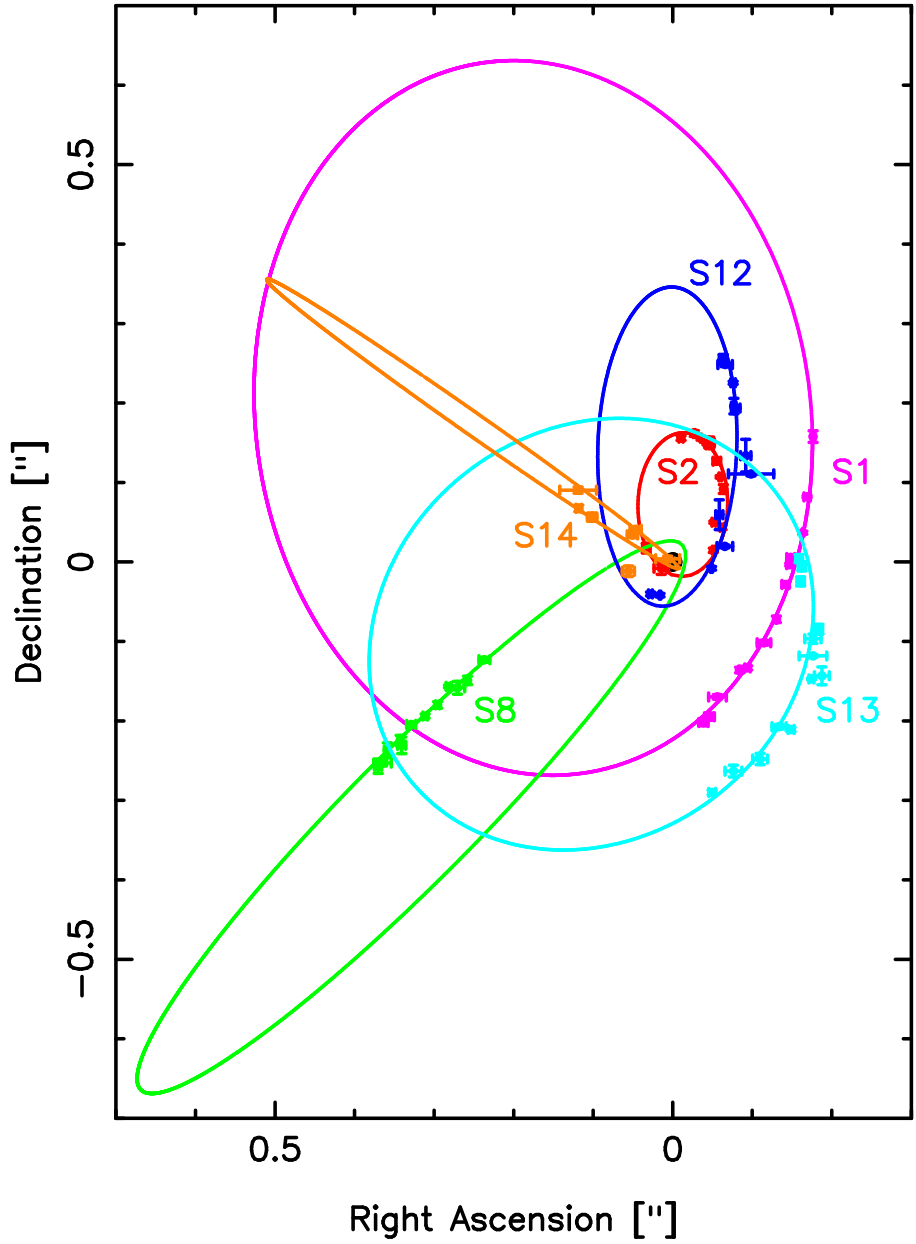}
\caption{The measured time dependent positions with their errors and
  the determined projected orbits of the stars S1, S2, S8, S12, S13,
  and S14. The corresponding orbital parameters are listed in
  Table~\ref{orbits}.
  \label{trajectories}}\end{figure}

\begin{figure}
\includegraphics[width=\textwidth,angle=270]{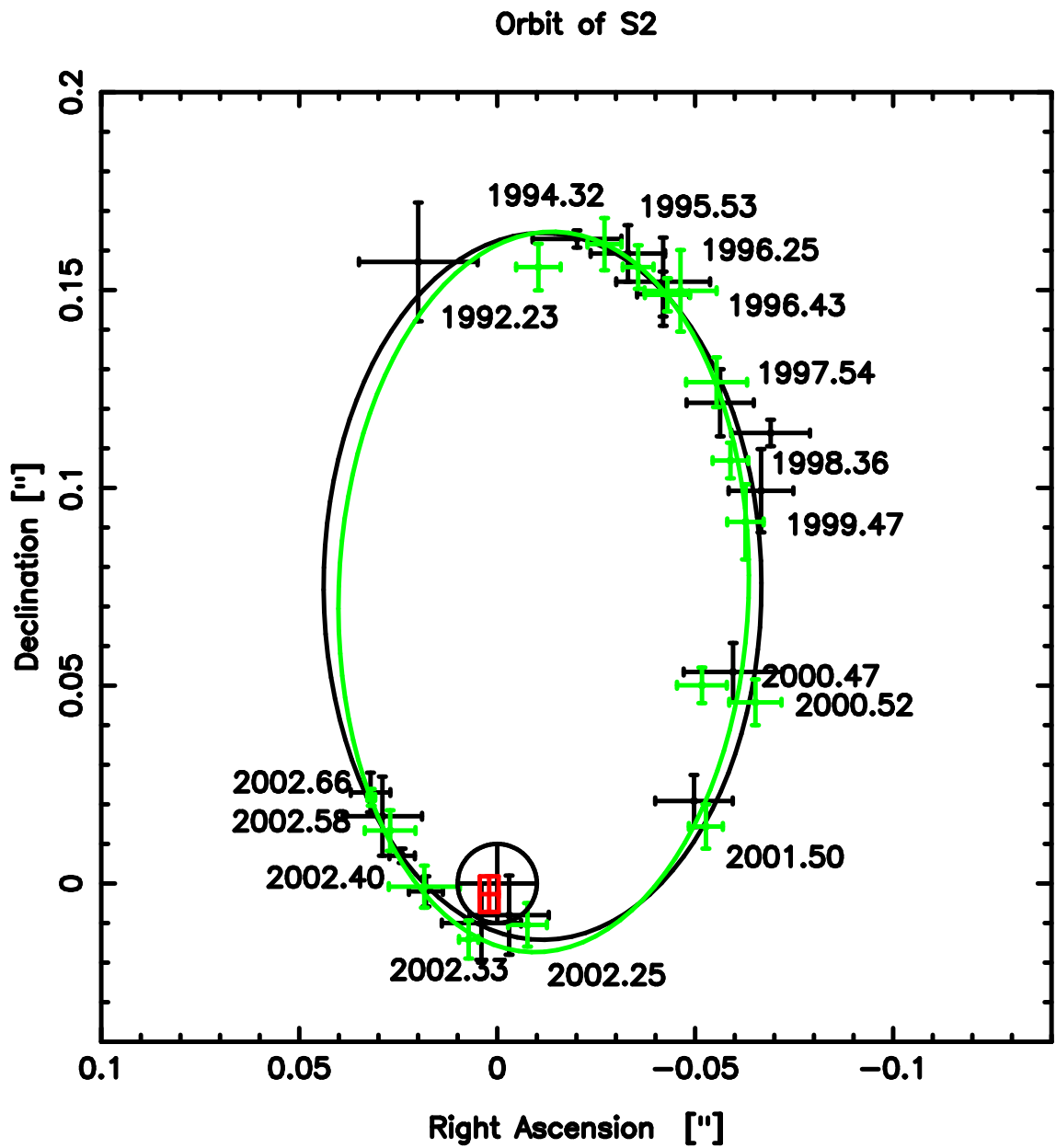}
\caption{Comparison of the measured positions and orbital fits for S2
  between this work (light grey) and \citet{schoedel02} (black).  The
  epoch labels refer to the positions from the present paper. The
  radio position of Sgr~A* is marked by a black cross and a 10~mas
  error circle. The rectangular box within this circle marks the
  position of the focus of the orbit and its errors as determined in
  the present work.\label{comporb}}
\end{figure}

\begin{figure}
\includegraphics[height=\textwidth,angle=270]{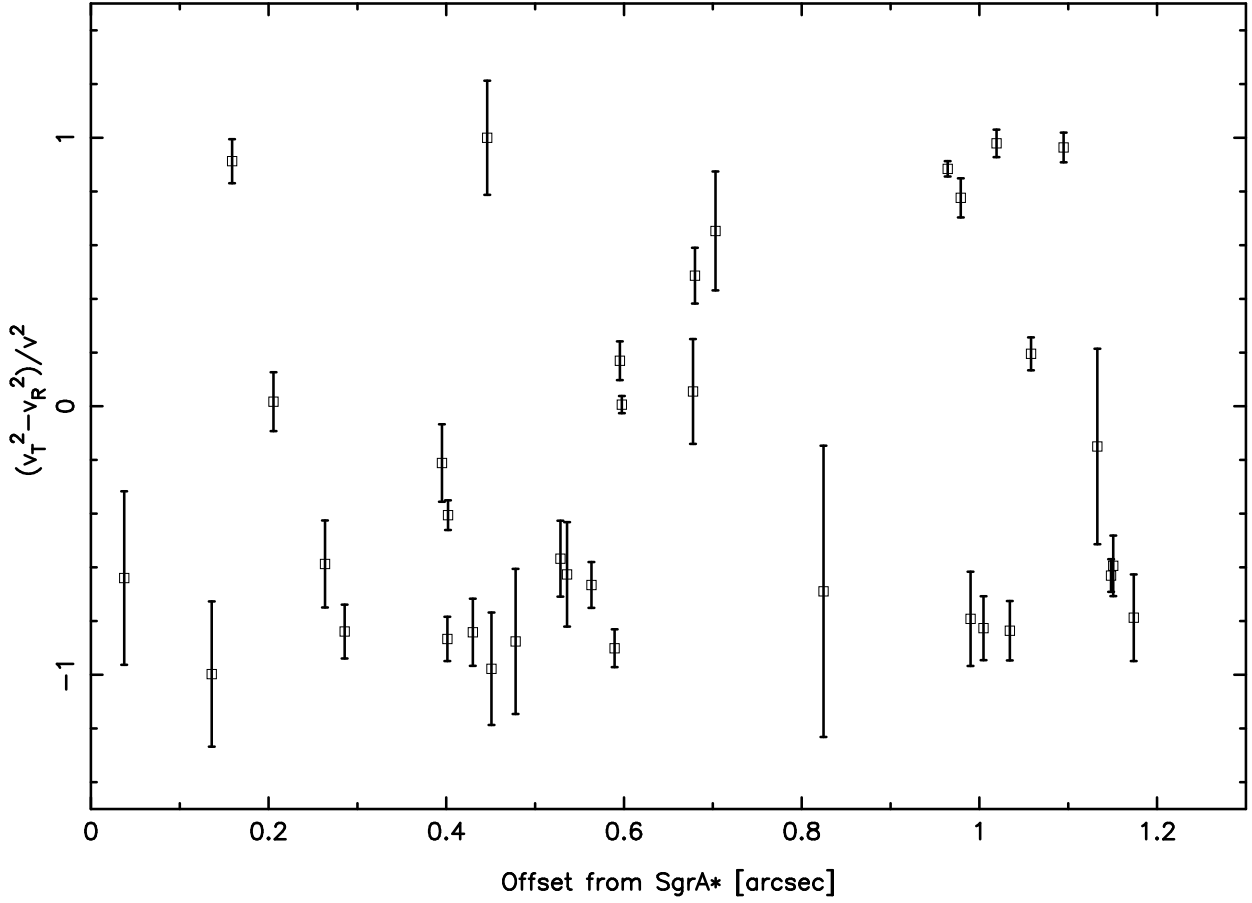}
\caption{The anisotropy parameter $\gamma_{TR}=(v_{T}^2-v_{R}^2)/v^2$
  plotted against the projected distance from Sgr~A* (epoch 2002.7)
  for the stars in Table~\ref{list} with the proper motions determined
  on the base of the entire data set.  $v$ is the proper motion velocity
  and $v_{T}$ and $v_{R}$ are its projected radial and tangential
  components. The majority of the stars appear to be on radial
  orbits.
  \label{aniso}}
\end{figure}

\begin{figure}
\includegraphics[width=12cm]{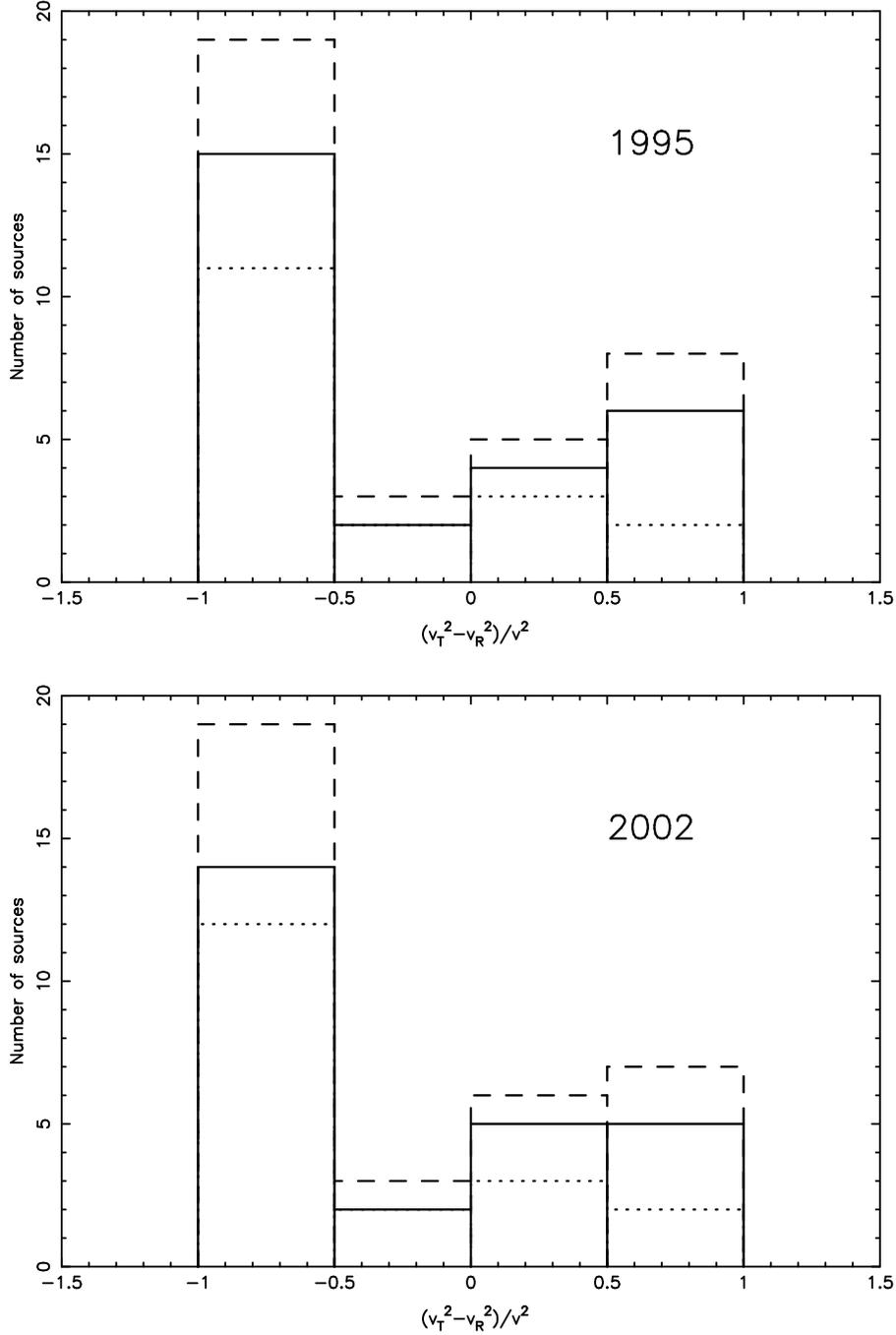}
\caption{Histogram of the anisotropy parameter
  $\gamma_{TR}=(v_{T}^2-v_{R}^2)/v^2$.  $v$ is the proper motion velocity and
  $v_{T}$ and $v_{R}$ are its projected radial and tangential
  components. Upper panel:  1995.5; lower panel: 2002.7. Dotted
  lines: Stars at projected distances $<0.6''$ from Sgr~A*; solid lines:
  stars at projected distances $< 1''$ ; dashed lines: stars at projected
  distances $< 1.2''$.
  \label{anisohist}}
\end{figure}

\begin{figure}
\includegraphics[height=\textwidth,angle=270]{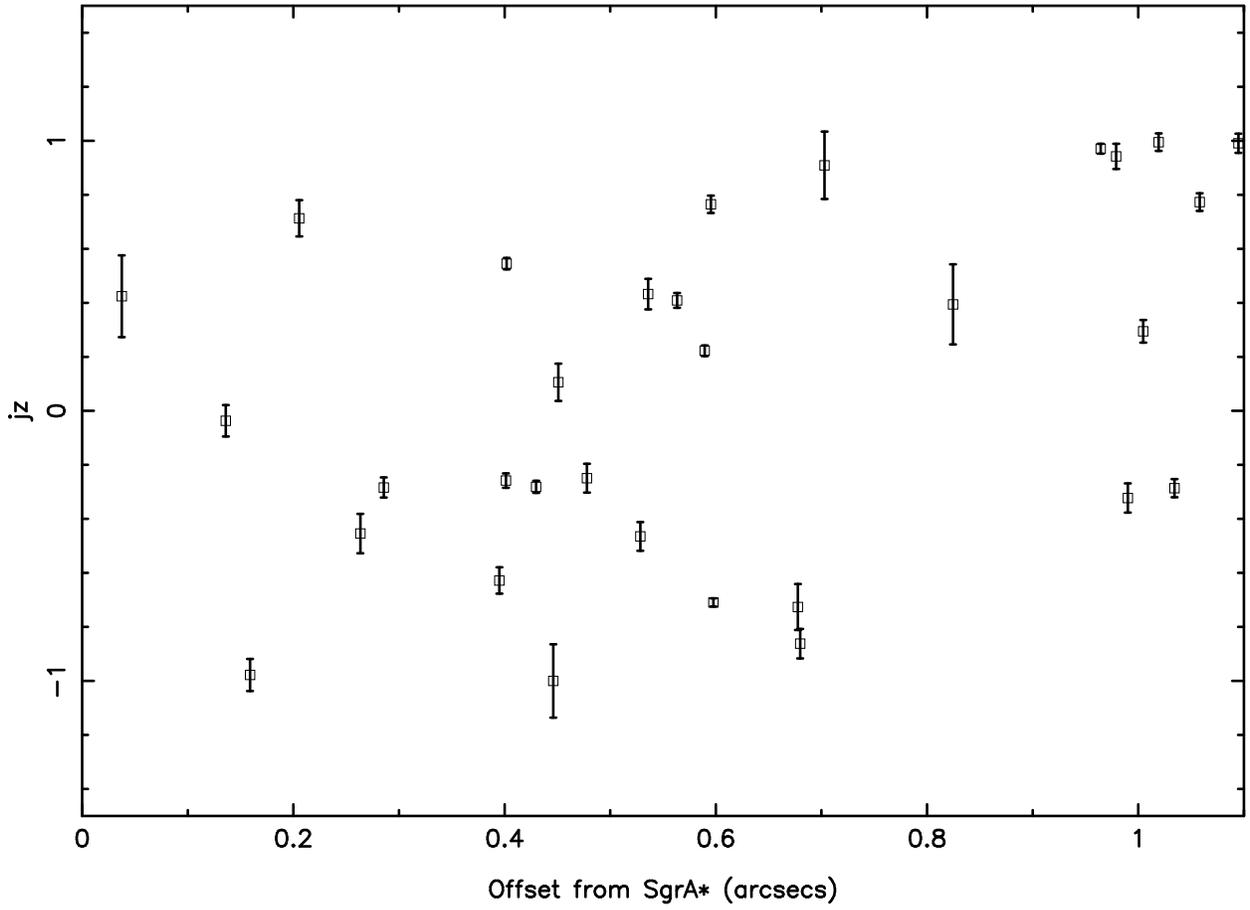}
\caption{The normalized angular momentum $J_{\mathrm{z}}/J_{\mathrm{z,max}} = (xv_y -
  yv_x)/pv_p$ in the 2002.7 epoch for the stars with proper motions based on the entire ten year data set.
  \label{angular}}
\end{figure}

\begin{figure}
\epsscale{0.8}
%\plotone{f11.eps}
\includegraphics[height=\textwidth,angle=270]{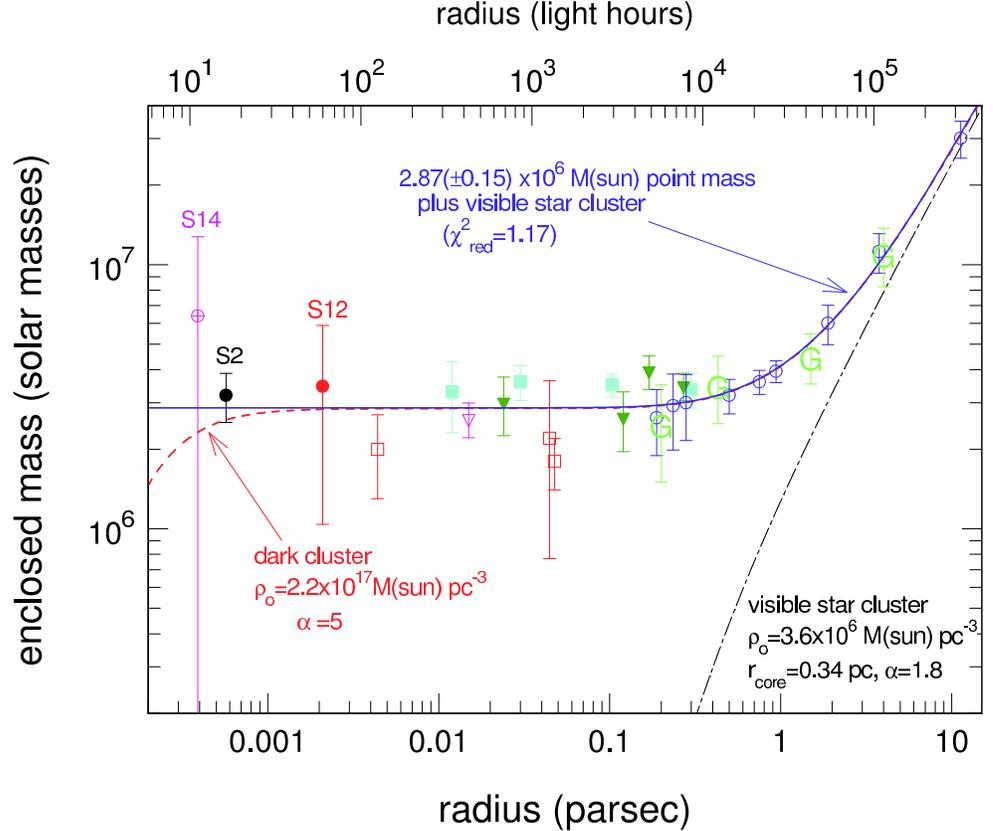}
\caption{\footnotesize Mass distribution in the Galactic Center
assuming an 8 kpc distance \citep{reid93}. The circles at the
shortest projected distances denote the masses derived from the orbits
of S2, S12, and S14. Filled down pointing triangles denote
Leonard-Merritt projected mass estimators from the present work (at
0.025~pc) and from a new NTT proper motion data set by \citet{ott03},
separating late and early type stars, and correcting for the volume
bias determined from Monte Carlo modeling of theoretical clusters and
assuming a central density profile with a power-law slope of
$\alpha=1.37$ \citep{genzel03}. An open down-pointing triangle denotes
the Bahcall-Tremaine mass estimate obtained from Keck proper motions
(Ghez et al. 1998; we multiplied their error bar by a factor of 2 in
order to take possible systematic errors into account). Filled
rectangles are mass estimates from a parameterized Jeans-equation
model, including effects of anisotropy and differentiating between
late and early type stars \citep{genzel00}. Open circles are mass
estimates from a parameterized Jeans-equation model of the radial
velocities of late type stars, assuming isotropy
\citep{genzel96}. Open rectangles denote mass estimates from a
non-parametric, maximum likelihood model, assuming isotropy and
combining late and early type stars \citep{chakrasaha01}. The
different statistical estimates (in part using the same or similar
data) agree within their uncertainties but the variations show the
sensitivity to the input assumptions. In contrast, the orbital
technique for S2, S12, and S14 is much simpler and less affected by the
assumptions. Letter ``G'' points denote mass estimates obtained from
Doppler motions of gas \citep{genzeltownes87}. The straight continuous
curve is the overall best fit model to all data. It is the sum of a
$2.87\pm0.15\times10^{6}$ M$_{\odot}$ point mass, plus a stellar
cluster of central density $3.6\times10^{6}$M$_{\odot}$pc$^{-3}$, core
radius 0.34 pc and power-law index $\alpha=1.8$. The grey long
dash-short dash curve shows the same stellar cluster separately, but
for an infinitely small core. The dashed curve is the
sum of the visible cluster, plus a Plummer model of a hypothetical
very compact (core radius $\sim$0.00019~pc) dark cluster of central
density $2.2\times10^{17}$M$_{\odot}$pc$^{-3}$.
\label{encmass}}
\end{figure}

\begin{figure}
\includegraphics[height=\textwidth,angle=270]{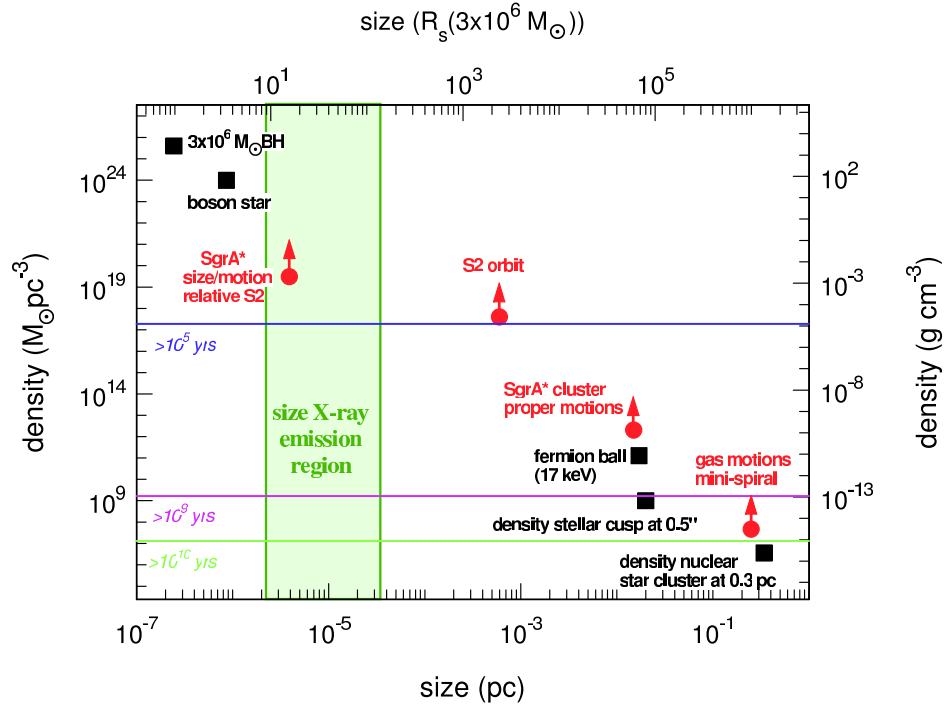}
\caption{Constraints on the nature of the observed dark mass in the
  Galactic Center. The horizontal axis is the size (bottom in parsec,
  top in Schwarzschild radii), and the vertical axis the density
  (left: M$_{\odot}$pc$^{-3}$, right: g~cm$^{-3}$). Filled circles
  denote the various limits on the size/density of the dark mass
  discussed in this paper. In addition the grey shaded area marks the
  constraints on the size of the variable X-ray emission from
  \citet{baganoff01}. Large filled squares mark the location of
  different physical objects, including the visible star cluster and
  its central cusp, the heavy neutrino, fermion ball of
  \citet{tsiviol98}, the boson star of \citet{torres00} and, in the
  top left, the position of a $\sim3\times10^{6}$M$_{\odot}$ black
  hole. In addition three thin horizontal lines mark the lifetimes of
  hypothetical dark clusters of astrophysical objects, e.g. neutron
  stars, white dwarfs, or stellar black holes \citep{maoz98}. The
  tightest constraints on the mass density come from the proper motion
  of Sgr~A* compared to the velocities in the surrounding star cluster
  if one assumes additionally that the size of the dark mass is given
  by the millimeter radio measurements of Sgr~A*. The now available
  measurements exclude all configurations but those of a black hole
  and a boson star.
\label{darkmass}}
\end{figure}

\begin{figure}
\includegraphics[height=\textwidth,angle=270]{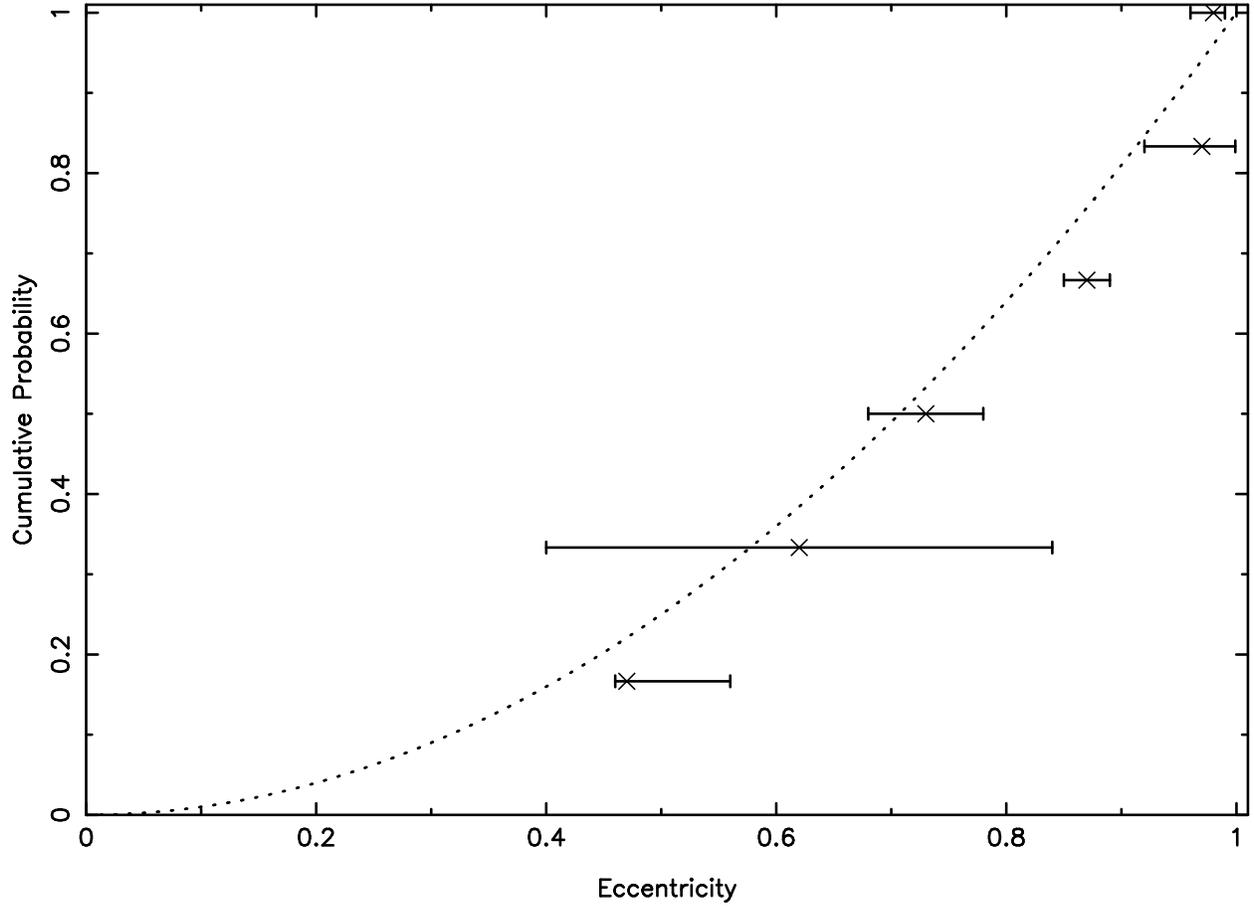}
\caption{Dotted line: Cumulative probability distribution function of
  the eccentricities of test particles orbiting a point mass in a
  spherical, isotropic system. The crosses and their error bars mark the
  eccentricities measured for the 6 stars S1, S2, S8, S12, S13, and
  S14.
  \label{epdf}}
\end{figure}

\begin{figure}
\includegraphics[height=\textwidth,angle=270]{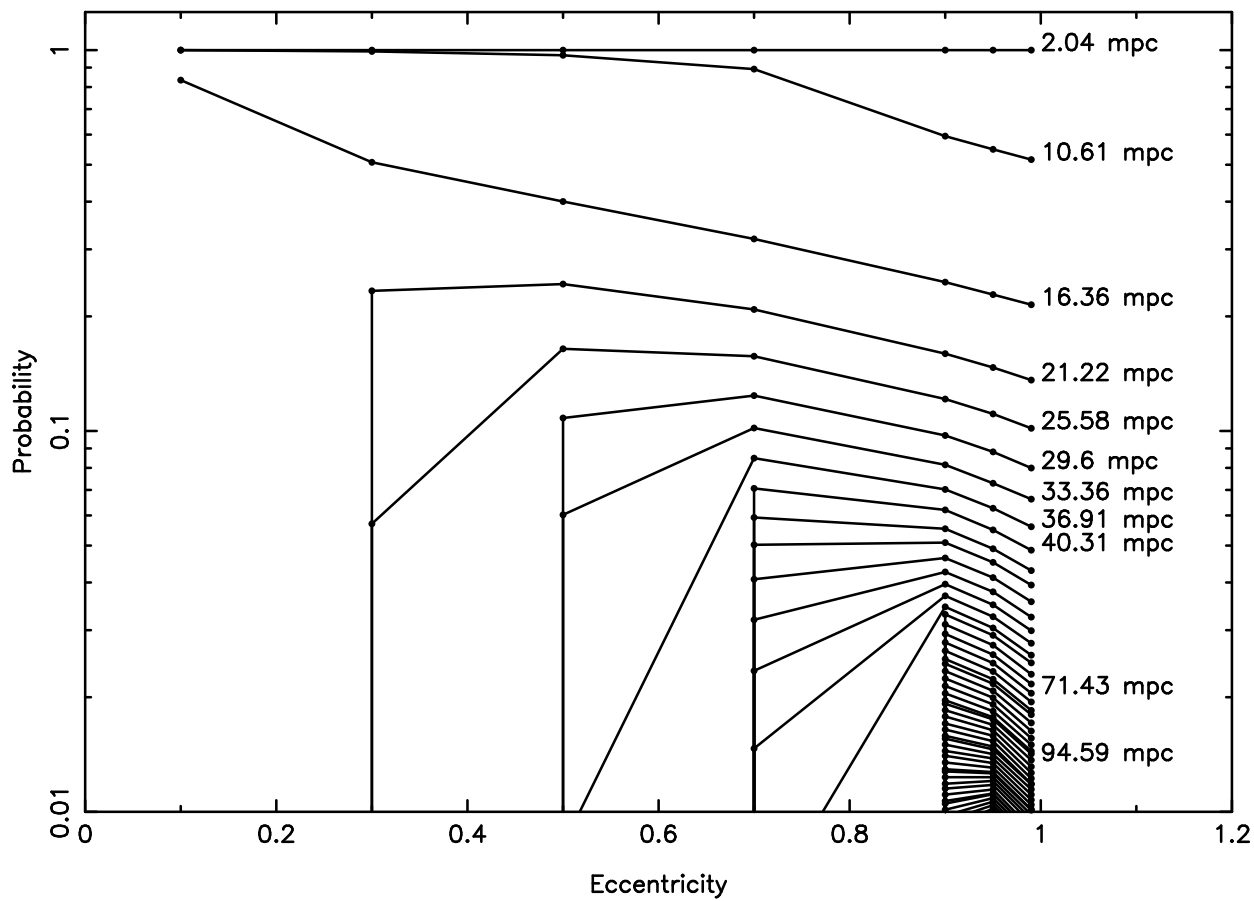}
\caption{Probability of detecting a keplerian orbit with a given
  eccentricity $e$ and semi-major axis $a$. The distribution of
  semimajor axes corresponds to the case of an isotropic stellar
  power-law cusp surrounding Sgr~A*. A stellar density law of
  $n_{*}\propto r^{-3/2}$ has been assumed \citep{genzel03}. Values of
  the semi-major axes are given in the labels to the right of the
  curves. See text for explanations concerning the model leading to
  the respective probabilities. Curves for semimajor axes
  $>100\quad\mathrm{mpc}$ were not plotted in order to avoid overcrowding
  the plot. \label{biasplot}}
\end{figure}

\begin{figure}
\includegraphics[height=\textwidth,angle=270]{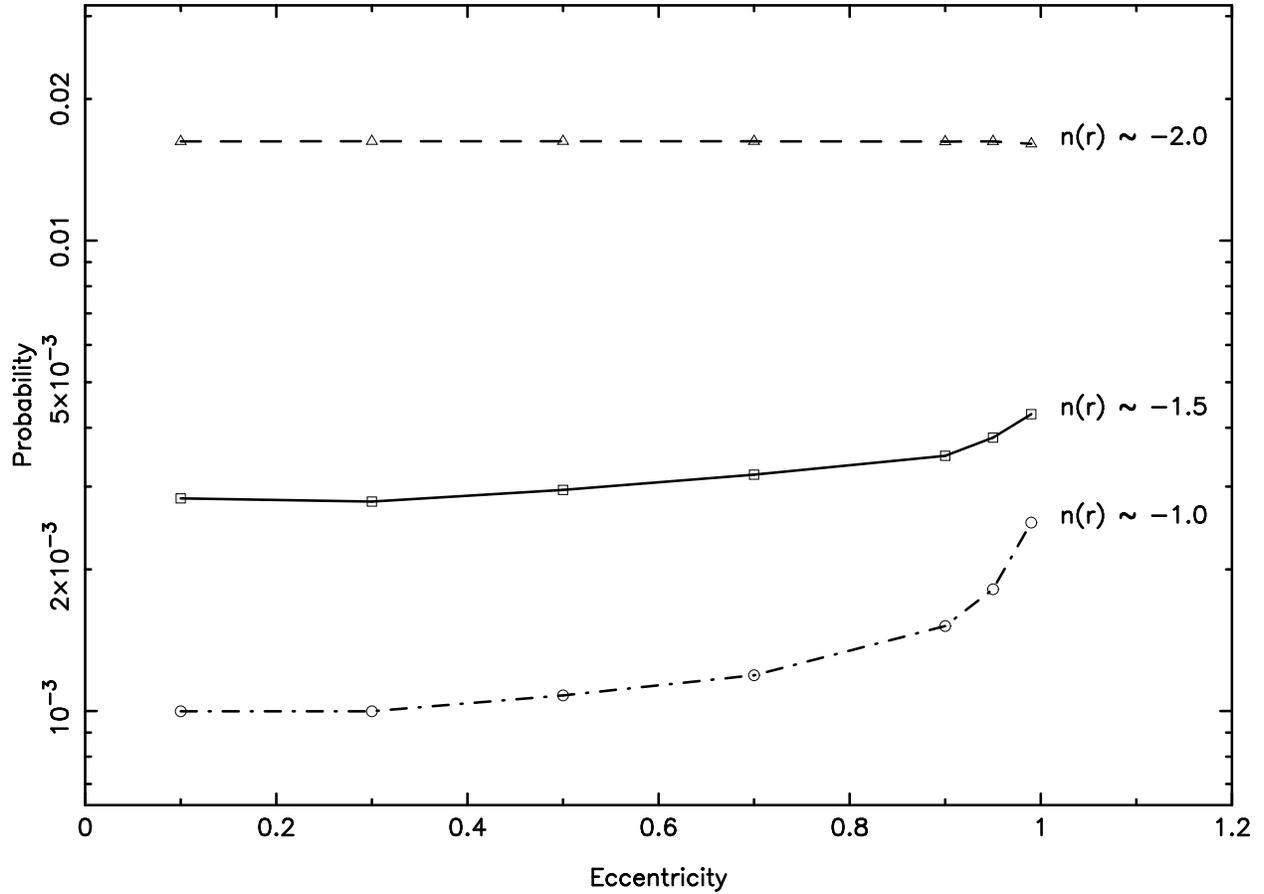}
\caption{Overall detection bias of orbits with a given eccentricity
  assuming a power-law stellar cusp at the center of the Milky Way.
  The exponents of the cusps are plotted beside the corresponding
  curves. The probabilities were obtained after averaging over the
  detection probabilities for given distributions of semimajor axes
  such as shown in Figure~\ref{biasplot}. The overall low
  probabilities result from the fact that most stars have large
  semimajor axes.
  \label{ebias}}
\end{figure}

\begin{figure}
\plotone{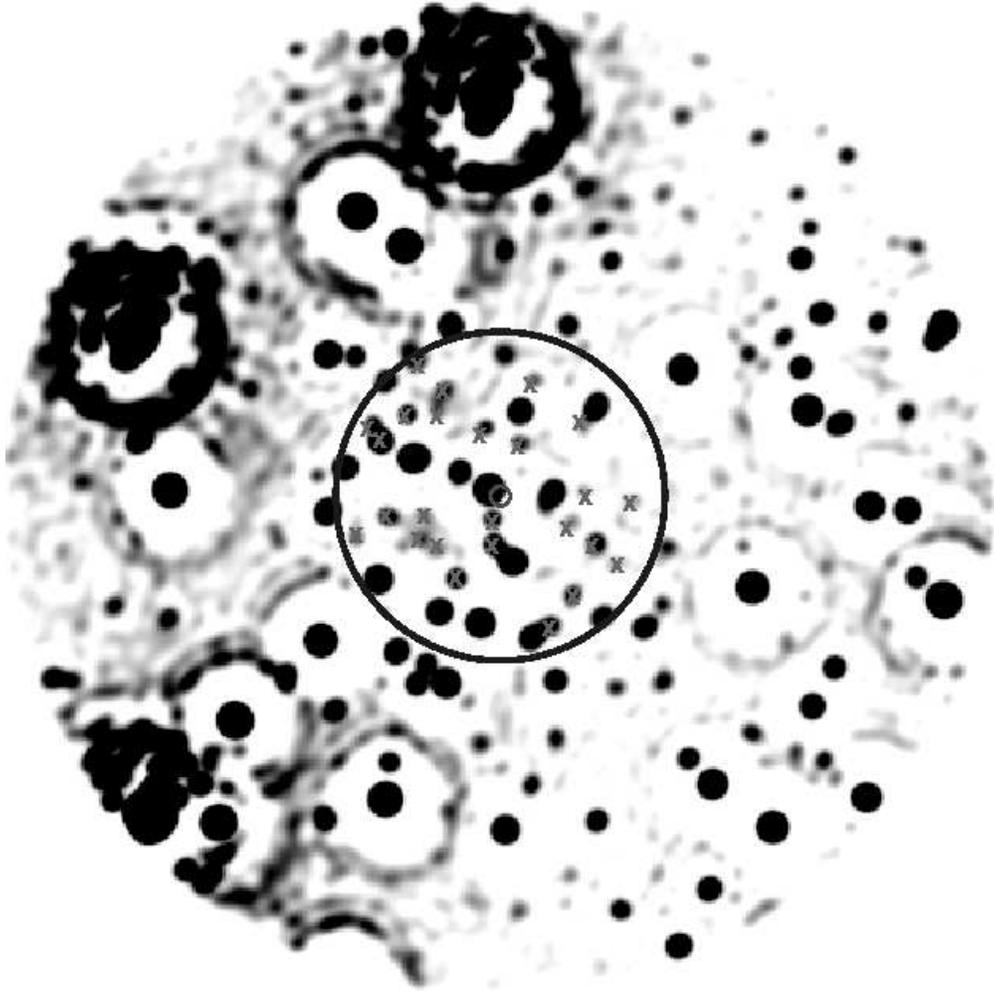}
\caption{ H-band map of the region $\leq1.5''$ around Sgr~A*, which is
  marked by a  30~mas radius circle. A 1500~s exposure
  shift-and-add image made with CONICA/NAOS in August 2002 was
  deconvolved with the Lucy-Richardson algorithm and smoothed to a
  FWHM of $\sim$40~mas. Rings around brighter sources are artifacts of
  the deconvolution algorithm. The brightest sources are
  saturated. The large circle marks the region wthin $0.5''$ of
  Sgr~A*. Some stars in that area with unknown or poorly known proper
  motions are marked with crosses. \label{gcnaco}}
\end{figure}

%Table of stellar proper motions

\clearpage

\begin{deluxetable}{lcccccccccccc}
\tabletypesize{\scriptsize}
\rotate 
\tablecaption{List of stars near
Sgr~A*.\label{list}} 
\tablewidth{0pt}

\tablehead{
\colhead{ID} & \colhead{Name} & \colhead{R} & \colhead{$M_{K}$} & \colhead{$dM_{K}$} & \colhead{$\Delta$R.A.} & \colhead{d$\Delta$R.A.} & \colhead{$\Delta$DEC} & \colhead{d$\Delta$DEC} & \colhead{$V_{R.A.}$} & \colhead{$dV_{R.A.}$} & \colhead{$V_{DEC}$} & \colhead{$dV_{DEC}$} \\ 
  &  & \colhead{(arcsec)} & \colhead{(mag)} & \colhead{(mag)} & \colhead{(arcsec)} & \colhead{(arcsec)} & \colhead{(arcsec)} & \colhead{(arcsec)} & \colhead{(km~s$^{-1}$)} & \colhead{(km~s$^{-1}$)} & \colhead{(km~s$^{-1}$)} & \colhead{(km~s$^{-1}$)} \\ 
\\
}

\startdata

1 & S2$^{*}$, S0-2  & 0.037 & 13.9 & 0.1 & 0.033 & 0.005 & 0.017 & 0.006 & 2913 & 166 & 3834 & 42 \\
2 & S14$^{*}$, S0-16  & 0.136 & 15.7 & 0.2 & 0.118 & 0.004 & 0.068 & 0.004 & 2106 & 190 & 1103 & 86 \\
3 & S13$^{*}$, S0-20 & 0.159 & 15.8 & 0.3 & -0.159 & 0.004 & 0.004 & 0.005 & 359 & 91 & 1483 & 48 \\
4 & S1$^{*}$, S0-1  & 0.206 & 14.7 & 0.1 & -0.039 & 0.006 & -0.202 & 0.004 & 801 & 19 & -1183 & 39 \\
5 & S12$^{*}$, S0-19 & 0.263 & 15.5 & 0.2 & -0.064 & 0.005 & 0.256 & 0.005 & 255 & 76 & 1098 & 52 \\
6 & S4, S0-3  & 0.286 & 14.4 & 0.1 & 0.262 & 0.004 & 0.113 & 0.008 & 623 & 17 & 74 & 14 \\
7$\dagger$ & \nodata & 0.288 & 16.5 & 0.3 & 0.129 & 0.006 & -0.257 & 0.007 & 984 & 187 & -207 & 207 \\
8$\dagger$ & \nodata  & 0.328 & 16.9 & 0.3 & -0.295 & 0.004 & -0.144 & 0.002 & -633 & 82 & 397 & 51 \\
9$\dagger$ & \nodata  & 0.371 & 16.4 & 0.3 & 0.281 & 0.007 & 0.242 & 0.003 & -237 & 135 & -246 & 93 \\
10$\dagger$ & \nodata  & 0.375 & 17.3 & 0.4 & -0.219 & 0.003 & -0.305 & 0.002 & -213 & 74 & 225 & 98 \\
11 & S10, S0-6  & 0.395 & 14.2 & 0.1 & 0.059 & 0.009 & -0.391 & 0.009 & -210 & 9 & 193 & 15 \\
12$\dagger$ & \nodata  & 0.399 & 15.2 & 0.1 & 0.362 & 0.010 & 0.168 & 0.007 & -12 & 180 & 330 & 185 \\
13 & S9, S0-5  & 0.401 & 15.1 & 0.1 & 0.184 & 0.004 & -0.356 & 0.005 & 109 & 12 & -499 & 11 \\
14 & , S0-8  & 0.402 & 15.8 & 0.3 & -0.296 & 0.002 & 0.272 & 0.005 & 121 & 11 & -471 & 6 \\
15 & \nodata  & 0.43 & 17.0 & 0.4 & -0.016 & 0.002 & 0.43 & 0.005 & 83 & 6 & 326 & 12 \\
16 & \nodata  & 0.446 & 15.4 & 0.2 & -0.1 & 0.007 & -0.435 & 0.006 & -105 & 10 & 23 & 12 \\
17 & S8$^{*}$, S0-4  & 0.451 & 14.5 & 0.1 & 0.37 & 0.004 & -0.257 & 0.009 & 536 & 40 & -569 & 36 \\
18 & S6, S0-7  & 0.478 & 15.4 & 0.2 & 0.47 & 0.008 & 0.085 & 0.007 & 295 & 23 & -21 & 14 \\
19$\dagger$ & \nodata & 0.482 & 16.6 & 0.3 & -0.313 & 0.01 & -0.366 & 0.003 & 152 & 190 & 175 & 210 \\
20 & S7, S0-11  & 0.528 & 15.3 & 0.2 & 0.526 & 0.005 & -0.048 & 0.006 & -225 & 10 & -93 & 12 \\
21 & \nodata  & 0.536 & 15.6 & 0.2 & 0.148 & 0.003 & 0.515 & 0.003 & -29 & 9 & 174 & 10 \\
22 & \nodata  & 0.563 & 16.7 & 0.3 & -0.21 & 0.004 & 0.523 & 0.004 & -11 & 7 & -290 & 7 \\
23 & \nodata  & 0.589 & 15.7 & 0.2 & -0.169 & 0.007 & -0.565 & 0.007 & 244 & 7 & 431 & 9 \\
24 & S11, S0-9  & 0.595 & 14.4 & 0.1 & 0.164 & 0.009 & -0.572 & 0.005 & 371 & 9 & -167 & 12 \\
25 & \nodata  & 0.598 & 16.2 & 0.2 & -0.442 & 0.003 & -0.402 & 0.004 & -774 & 9 & 40 & 8 \\
26 & \nodata  & 0.678 & 15.1 & 0.1 & 0.525 & 0.006 & 0.429 & 0.005 & 153 & 11 & -20 & 10 \\
27 & W6, S0-12  & 0.68 & 14.4 & 0.1 & -0.562 & 0.003 & 0.383 & 0.004 & 13 & 9 & 190 & 8 \\
28 & , S0-13  & 0.703 & 13.4 & 0.1 & 0.546 & 0.006 & -0.443 & 0.009 & 34 & 13 & 132 & 12 \\
29$\dagger$ & \nodata & 0.752 & 17.0 & 0.4 & -0.496 & 0.004 & -0.565 & 0.003 & -99 & 64 & 590 & 110 \\
30 & W9, S0-14  & 0.825 & 13.8 & 0.1 & -0.775 & 0.002 & -0.28 & 0.004 & 34 & 6 & -2 & 4 \\
31$\dagger$ & \nodata  & 0.887 & 17.0 & 0.3 & -0.094 & 0.005 & -0.882 & 0.005 & -108 & 124 & 209 & 167 \\
32 & , S1-3  & 0.964 & 12.3 & 0.1 & 0.429 & 0.006 & 0.864 & 0.003 & -518 & 6 & 115 & 9 \\
33 & W5, S0-15  & 0.979 & 13.7 & 0.1 & -0.943 & 0.004 & 0.261 & 0.008 & -262 & 11 & -374 & 16 \\
34 & , S1-5  & 0.99 & 12.6 & 0.1 & 0.348 & 0.008 & -0.927 & 0.005 & -164 & 13 & 199 & 13 \\
35$\dagger$ & \nodata  & 1.003 & 15.6 & 0.1 & -0.924 & 0.006 & 0.392 & 0.004 & 103 & 162 & -140 & 144 \\
36 & , S1-1  & 1.005 & 13.2 & 0.1 & 1.005 & 0.005 & 0.017 & 0.004 & 223 & 8 & 73 & 9 \\
37 & , S1-2  & 1.019 & 14.9 & 0.1 & -0.019 & 0.004 & -1.019 & 0.004 & 453 & 10 & -55 & 13 \\
38 & \nodata  & 1.034 & 16.1 & 0.1 & -0.296 & 0.003 & -0.991 & 0.004 & -193 & 11 & -294 & 11 \\
39 & , S1-4  & 1.058 & 12.6 & 0.1 & 0.806 & 0.004 & -0.685 & 0.005 & 441 & 8 & 80 & 15 \\
40$\dagger$ & \nodata & 1.065 & 15.8 & 0.3 & -1.042 & 0.004 & 0.219 & 0.003 & 82 & 135 & -280 & 111 \\
41 & , S1-8  & 1.095 & 14.2 & 0.1 & -0.652 & 0.003 & -0.88 & 0.007 & 296 & 8 & -162 & 9 \\
42 & \nodata  & 1.133 & 15.7 & 0.1 & -0.984 & 0.005 & 0.561 & 0.007 & -102 & 13 & -20 & 14 \\
43$\dagger$ & \nodata  & 1.135 & 14.8 & 0.1 & -1.135 & 0.004 & -0.029 & 0.005 & 37 & 83 & -37 & 119 \\
44 & , S1-7  & 1.148 & 15.8 & 0.1 & -1.023 & 0.003 & -0.522 & 0.005 & -212 & 5 & -277 & 8 \\
45 & \nodata  & 1.151 & 15.3 & 0.1 & -0.956 & 0.003 & -0.641 & 0.006 & -236 & 6 & -29 & 13 \\
46 & , S1-6  & 1.174 & 15.6 & 0.1 & -0.923 & 0.007 & 0.725 & 0.008 & -309 & 15 & 107 & 15 \\

\tablecomments{ List of stars within $1.2''$ of Sgr~A*. Stars marked
    with an asterisk show accelerated movement. The velocities given
    for these stars are approximate velocities for the 2002.7 epoch,
    derived from a linear fit to a subset of the positional
    data. Positions and magnitudes are for the epoch 2002.66. An
    additional general $\sim$0.1 magnitude systematic error should be
    taken into account because of uncertainties in the fluxes of the
    calibration sources. Proper motions of stars marked with a
    $\dagger$ are based solely on the Gemini/NACO data subset as
    described in the text. Names behind a comma in column~2 refer to
    names assigned by \citet{ghez98}, except S0-16, S0-19, and S0-20,
    which refer to \citep{ghez03a}. In section~\ref{orbitsection} we
    estimate an additional systematic error of the order $0.003''$ on
    the positions at all epochs. That corresponds to an additional
    error of $\sim$20~km/s in the proper motion velocities.}

\enddata
\end{deluxetable}

\clearpage
\addtocounter{page}{-1}

% Table of stellar accelerations

\begin{deluxetable}{lcccccccccc}
\rotate \tablecaption{List of stars near Sgr~A* with measured average
accelerations from parabolic fits to sections of their
orbits. \label{accellist}} \tablewidth{0pt} \tablehead{ \colhead{ID} &
\colhead{Name} & \colhead{Epoch} & \colhead{$\Delta$R.A.}  &
\colhead{$\Delta$DEC} & \colhead{a} & \colhead{da} &
\colhead{$a_{R.A.}$} & \colhead{$da_{R.A.}$} & \colhead{$a_{DEC}$} &
\colhead{$da_{DEC}$} \\ & & \colhead{(yr)} & \colhead{(arcsec)} &
\colhead{(arcsec)} & \colhead{(mas~yr$^{-2}$)} &
\colhead{(mas~yr$^{-2}$)} & \colhead{(mas~yr$^{-2}$)} &
\colhead{(mas~yr$^{-2}$)} & \colhead{(mas~yr$^{-2}$)} &
\colhead{(mas~yr$^{-2}$)} \\ \\ } \startdata

1 & S2 & 1997.25 & -0.051 & 0.132 & 5.55 & 0.66 & 1.12 & 0.53 & -5.43 & 0.67\\
3 & S13 & 1998.38 & -0.179 & -0.135 & 5.70 & 0.30 & 5.69 & 0.30 & -0.12 & 0.21\\
4 & S1 & 1998.16 & -0.131 & -0.059 & 3.84 & 0.27 & 3.64 & 0.25 & 1.23 & 0.35 \\
5 & S12 & 2000.75 & -0.077 & 0.201 & 7.03 & 0.57 & 4.16 & 0.73 & -5.66 & 0.47\\
17 & S8 & 1998.16 & 0.307 & -0.187 & 1.37 & 0.46 & -1.24 & 0.47 & 0.58 & 0.46\\

\enddata
\end{deluxetable}

% Table of stellar accelerations

\begin{deluxetable}{lcccccc}
\tablecaption{Projected accelerations, projected masses and 
  inclination angles, determined from parabolic fits. \label{masslist}}
\tablewidth{0pt}
\tablehead{
\colhead{ID} & \colhead{Name} & \colhead{a} &
\colhead{da} & \colhead{M$_{proj}=M_{BH}*cos(i)^{3}$} & \colhead{i} & \colhead{di}  \\ 
  &  & \colhead{(mas~yr$^{-2}$)} & \colhead{(mas~yr$^{-2}$)} & \colhead{$10^{6}\times$M$_{\odot}$} & \colhead{deg} & \colhead{deg}  \\ 
}

\startdata
1  & S2 & 5.55 & 0.66 & 1.5 & 37 & 6 \\
3  & S13 & 5.70 & 0.30 & 4.48 & n.a. & n.a.\\
4  & S1 & 3.84 & 0.27 & 0.89 & 47 & 5 \\
5  & S12 & 7.03 & 0.57 & 4.34 & n.a. & n.a. \\
17 & S8 & 1.37 & 0.46 & 2.50 & 17 & 42 \\
\enddata

\end{deluxetable}

% Table of S2 parameters
% **********************

\begin{center}
\begin{table}
\caption{Orbital parameters of S2 as determined for normal (column~1)
  and equal weighting (column~2) of the measured positions. In the
  third column we list the orbital parameters of S2 according to
  \citet{ghez03}. Offsets are measured positive to the East and North.
  \label{compweight}}
\begin{tabular}{lccc}
Weighting & Normal & Equal & Ghez et al.\\
\hline
\hline
Offset R.A. (mas)                  & $2.0\pm2.4$      &  $1.4\pm1.3$  & $-2.7\pm1.9$ \\
Offset Decl. (mas)                 & $-2.7\pm4.5$      &  $-2.3\pm3.1$  & $-5.4\pm1.4$ \\
Central Mass ($10^{6}$M$_{\odot}$) & 3.31$\pm$0.67    &  3.07$\pm$0.72 & 4.07$\pm$0.68\\
Period (yr)                        & 15.73$\pm$0.74   & 15.31$\pm$0.71 & 15.78$\pm$0.82 \\
Pericenter Passage (yr)            & 2002.31$\pm$0.02 &  2002.31$\pm$0.04 & 2002.33$\pm$0.02 \\
Eccentricity                       & 0.87$\pm$0.02    &  0.87$\pm$0.02 & 0.87$\pm0.01$ \\
Angle of line of nodes (deg)       & 44.2$\pm$7.0     &  49.1$\pm$8.2 & 40.1$\pm$3.0 \\
Inclination (deg)                  & $\pm$45.7$\pm$2.6     &  $\pm$44.7$\pm$3.9 & -47.3$\pm$2.5 \\
Angle of node to pericenter (deg)  & 244.7.1$\pm$4.7  &  241.0$\pm$6.3 & 251.4$\pm$1.8 \\
Semi-major axis (mpc)              & 4.54$\pm$0.27    &  4.34$\pm$0.31 & 4.87$\pm$0.21 \\
Separation of pericenter (mpc)     & 0.59$\pm$0.10    &  0.56$\pm$0.10 & 0.61$\pm$0.05 \\
\hline
\end{tabular}
\end{table}
\end{center}

% Table of orbits
% ***************

\begin{center}
\begin{table}
\caption{Parameters and their formal uncertainties for the orbits of
S1, S2, S8, S12, S13 and S14. The position of the focus has been
fitted simultaneously in the case of S2. For the other stars, additional
astrometric errors of the parameters must be taken into account because
of the uncertainty in the position of the gravitational focus. These
errors are of the same order as the fitting errors. \label{orbits}}
\scriptsize{
\begin{tabular}{lcccccc}
 & S2  & S12 & S14 & S1 & S8 & S13\\
\hline
\hline
Central Mass ($10^{6}$M$_{\odot}$) & 3.3$\pm$0.7    & 3.5$\pm$1.8 &  6.4$\pm$7.6 & 2.6$^{2.1}_{3.5}$          & 1.9$^{1.6}_{3.1}$ & 2.3$^{2.2}_{2.9}$ \\
Period (yr)                        & 15.73$\pm$0.74   &  36.0$\pm$8.3   & 69.0$\pm$26.6  & 171$^{602}_{86}$           & 342$^{473}_{148}$ & 75$^{75}_{148}$ \\
Pericenter Passage (yr)            & 2002.31$\pm$0.02 & 1995.3$\pm$0.3 & 1999.9$\pm$0.2 & 1999.6$^{1999.1}_{2000.6}$  & 2330$^{2461}_{2134}$ & 2005.4$^{2005.4}_{2005.1}$  \\ 
Eccentricity                       & 0.87$\pm$0.02    & 0.73$\pm$0.13    & 0.97$\pm$0.05  & 0.62$^{0.84}_{0.40}$        & 0.98$^{0.99}_{0.96}$ &  0.47$^{0.46}_{0.56}$ \\
Angle of line of nodes (deg)       & 44.2$\pm$7.0     & 99.9$\pm$2.3     & 40.2$\pm$3.0   & 104$^{103}_{105}$           & 109$^{106}_{124}$ &  173$^{138}_{175}$\\
Inclination (deg)                  & 45.7$\pm$2.6     & 52.6$\pm$6.9     &  82.6$\pm$0.5  & 45$^{40}_{50}$              & 30$^{10}_{60}$ &  10$^{0}_{30}$ \\
Angle of node to pericenter (deg)  & 244.7.1$\pm$4.7    & 159.9$\pm$16.3  & 143.0$\pm$19.5 & 114$^{111}_{122}$           & 25$^{25}_{23}$ & 325$^{0}_{320}$ \\
Semi-major axis (mpc)              & 4.54$\pm$0.27    & 8.0$\pm$0.7    & 15.1$\pm$4.6   & 21$^{45}_{14}$              & 29$^{34}_{20}$ & 11$^{11}_{13}$ \\
Separation of pericenter (mpc)     & 0.59$\pm$0.10    & 2.2$\pm$0.4    & 0.39$\pm$0.78  & 8$^{7}_{9}$                 & 0.6$^{0.3}_{0.8}$ & 6$^{6}_{6}$ \\
\hline
\end{tabular}}
\end{table}
\end{center}

%Table of mass estimates
\begin{center}
\begin{table}
\caption{Table of enclosed mass estimates from proper motions of stars
  within $1.2''$ of Sgr~A*. We compiled four lists of stellar proper
  motions: 2 short lists including 35 proper motions from the entire
  SHARP/Gemini/NACO data set and 2 long lists (LL) including
  additionally the (larger error) proper motions of 11 sources solely
  based on the Gemini July 2000/NACO September 2002 images. For the 6
  stars with measured accelerated motion, we determined velocities at
  two different epochs each and included them in the short and long
  lists (SL1, SL2, LL1, LL2). $N$ is the number of stars in the
  respective list, $p$ is the projected distance from Sgr~A* (please
  refer to Bahcall \& Tremaine 1981 for the meaning of the expression
  $2/\pi/\langle1/p\rangle$), $M_{LM}$ is the Lonard-Merritt mass
  estimator, $\sigma_{R}$ is the projected radial and $\sigma_{T}$ the
  projected tangential velocity dispersion.
  \label{masses}}

\begin{tabular}{lcccccc}
List & N & $2/\pi/\langle1/p\rangle$ & $M_{LM}$ &  $\sigma_{R}$ & $\sigma_{T}/\sigma_{R}$ \\
 & & ($''$) & $1\times10^{6}$M$_{\odot}$ & $1\times10^{6}$M$_{\odot}$ & &  \\
\hline
\hline
SL1 & 35 & 0.58 & $3.7\pm0.9$ & $840\pm104$ & $0.66\pm0.12$ \\
SL2 & 35 & 0.67 & $3.7\pm0.9$ & $540\pm67$ & $0.78\pm0.12$ \\
LL1 & 46 & 0.61 & $3.3\pm0.7$ & $745\pm78$ & $0.68\pm0.10$ \\
LL2 & 46 & 0.68 & $3.1\pm0.7$ & $498\pm52$ & $0.98\pm0.14$ \\
\hline
\end{tabular}
\end{table}
\end{center}

%Table of most recent mass estimates
\begin{center}
\begin{table}
\caption{Different mass estimates for the central point mass (in
  $10^{6}$M$_{\odot}$).  Column~1: Mass estimate from the orbit of S2
  (this work). Column~2: Mass estimate from the orbit of S2 as
  determined by \citet{ghez03a}. Column~3: Average mass estimate from
  three stellar orbits by \citet{ghez03a}. Column~4: Average LM mass
  estimate from proper motions within $1.2''$ of SgrA* (this
  work). Column~5: Mass estimate from fit to all available
  measurements within $\sim1$~pc of Sgr~A* (see Figure~\ref{encmass},
  this work). Column~6: Like column~5, without the systematically low
  estimates (see section~\ref{bhmass}).
  \label{bestmasses}}

\begin{tabular}{cccccc}
M$_{S2,1}$ & M$_{S2,2}$  & M$_{3orbits}$ & M$_{LM}$ ($<1.2''$) & Fit$_{1}$ & Fit$_{2}$\\
\hline
\hline
$3.3\pm0.7$ & $4.1\pm0.6$ & $3.6\pm0.4$ & $3.4\pm0.5$ & $2.9\pm0.2$ &  $3.1\pm0.2$\\
\hline
\end{tabular}
\end{table}
\end{center}

\end{document}